\documentclass{ecai}  

\usepackage{graphicx}
\usepackage{latexsym}
\usepackage{times}
\usepackage{epsfig}
\usepackage{amsmath}
\usepackage{amssymb}
\usepackage{algorithm}
\usepackage{algorithmic}
\usepackage{booktabs}
\usepackage{multirow}
\usepackage{tikz}
\usepackage{xcolor}
\usepackage{mathtools}
\usepackage{adjustbox}
\usepackage{orcidlink}
\usepackage{caption}
\usepackage[switch]{lineno}
\newtheorem{assumption}[theorem]{Assumption}
\newenvironment{proof}[1][Proof]{%
  \par\medskip\noindent\textbf{#1.}\ }%
  {\hfill\(\square\)\par\medskip}

\begin{document}

\begin{frontmatter}

\title{HMAE: Self-Supervised Few-Shot Learning for Quantum Spin Systems}

\author[A]{\fnms{Ibne Farabi}~\snm{Shihab}\orcidlink{0000-0003-1624-9954}\thanks{*Equal contribution. Corresponding author. Email: ishihab@iastate.edu.}}
\author[A]{\fnms{Sanjeda}~\snm{Akter}\orcidlink{0009-0007-8276-3878}\thanks{*Equal contribution.}}
\author[B]{\fnms{Anuj}~\snm{Sharma}\orcidlink{0000-0001-5929-5120}}

\address[A]{Department of Computer Science, Iowa State University, Ames, Iowa, USA}
\address[B]{Department of Civil, Construction and Environmental Engineering, Iowa State University, Ames, Iowa, USA}

\begin{abstract}
Quantum machine learning for spin and molecular systems faces critical challenges of scarce labeled data and computationally expensive simulations. To address these limitations, we introduce Hamiltonian-Masked Autoencoding (HMAE), a novel self-supervised framework that pre-trains transformers on unlabeled quantum Hamiltonians, enabling efficient few-shot transfer learning. Unlike random masking approaches, HMAE employs a physics-informed strategy based on quantum information theory to selectively mask Hamiltonian terms based on their physical significance. Experiments on 12,500 quantum Hamiltonians (60\% real-world, 40\% synthetic) demonstrate that HMAE achieves 85.3\% $\pm$ 1.5\% accuracy in phase classification and 0.15 $\pm$ 0.02 eV MAE in ground state energy prediction with merely 10 labeled examples—a statistically significant improvement (p < 0.01) over classical graph neural networks (78.1\% $\pm$ 2.1\%) and quantum neural networks (76.8\% $\pm$ 2.3\%). Our method’s primary advantage is exceptional sample efficiency—reducing required labeled examples by 3-5× compared to baseline methods—though we emphasize that ground truth values for fine-tuning and evaluation still require exact diagonalization or tensor networks. We explicitly acknowledge that our current approach is limited to small quantum systems (specifically limited to 12 qubits during training, with limited extension to 16-20 qubits in testing) and that, while promising within this regime, this size restriction prevents immediate application to larger systems of practical interest in materials science and quantum chemistry.
\end{abstract}

\end{frontmatter}
\section{Introduction}
Quantum systems like spin lattices and small molecules offer vast potential for materials innovation but are difficult to study due to exponential Hilbert space growth \cite{nielsen2010quantum, sachdev2011quantum}. Current approaches use computationally intensive simulations scaling as $O(2^n)$ for $n$ qubits \cite{troyer2005computational, daley2022practical}, limiting exploration of complex quantum materials and molecules.

Machine learning has transformed scientific fields through pre-training and transfer learning \cite{devlin2018bert, he2022masked, brown2020language}, yet quantum systems present unique challenges. Their complex-valued, high-dimensional nature hinders application of conventional machine learning techniques \cite{cerezo2022challenges}. Current quantum machine learning approaches require substantial labeled data \cite{carrasquilla2017machine, beach2019qucumber} and lack transfer capabilities, while quantum mechanics imposes constraints that generic models often fail to respect.

A quantum Hamiltonian $H$ describes a system's energy and governs its behavior. For spin systems like the Heisenberg model, interactions between quantum spins can be represented using Pauli matrices: $H = J \sum_{<i,j>} \vec{\sigma}_i \cdot \vec{\sigma}_j$. In molecular systems, Hamiltonians describe electron interactions using creation and annihilation operators. Despite their exponential scaling with system size, Hamiltonians exhibit inherent structure through local interactions, symmetries, and conservation laws that can enable more efficient learning.

We introduce Hamiltonian-Masked Autoencoding (HMAE), a self-supervised framework pre-training transformers on unlabeled quantum Hamiltonians for efficient few-shot transfer learning. HMAE employs a physics-informed masking strategy based on quantum information theory to selectively mask Hamiltonian terms by physical significance. Our key insight is that quantum Hamiltonians' physical structure provides natural self-supervision. By reconstructing physically significant terms, our model learns representations respecting quantum symmetries and conservation laws.

We demonstrate HMAE's effectiveness through experiments on 12,500 quantum Hamiltonians (60\% real-world, 40\% synthetic). Our approach significantly improves few-shot learning for phase classification and ground state energy prediction compared to graph neural networks and quantum neural networks. Concurrent approaches to quantum self-supervised learning have emerged independently, highlighting this research direction's significance. Our approach uniquely combines physics-informed masking grounded in quantum information theory, superior cross-platform transferability, and theoretical foundations connecting to quantum mutual information principles.

The paper is organized as follows: Section \ref{sec:related_work} discusses related work, Section \ref{sec:Quantum_self} details our quantum self-supervised learning framework, Section \ref{sec:Experiment} presents experimental results, and Section \ref{sec:con} provides conclusions and discusses limitations and future directions.
\section{Related Work}
\label{sec:related_work}

Machine learning for quantum systems includes supervised approaches for tasks like ground state prediction \cite{carrasquilla2017machine} and phase classification \cite{van2017learning}, quantum neural networks (QNNs) implemented on quantum hardware \cite{mitarai2018quantum, benedetti2019parameterized}, and physics-informed neural networks (PINNs) that incorporate physical constraints \cite{karniadakis2021physics, raissi2019physics}. Classical ML has achieved notable successes in quantum applications, such as identifying phase transitions in many-body systems \cite{carrasquilla2017machine} and predicting molecular properties \cite{unke2021machine}. Despite these successes, existing approaches typically require substantial labeled data, struggle with transferability across quantum domains, and many require error-prone quantum hardware. QNNs face the ``barren plateau'' problem \cite{mcclean2018barren}, where gradients vanish exponentially with system size, limiting their practical application.

Self-supervised methods have transformed AI through masked language modeling \cite{devlin2018bert}, contrastive learning \cite{chen2020simple}, and masked autoencoders \cite{he2022masked}. However, these advances have not been successfully adapted for quantum systems due to the complex-valued nature of quantum data, lack of natural tokenization strategies, and the need to respect physical constraints like unitarity and Hermiticity. Previous attempts at applying transformers to quantum data have either ignored these constraints or limited themselves to simple quantum systems with artificial structure.

Various approaches for quantum representation include tensor networks \cite{stoudenmire2016supervised}, graph neural networks adapted for quantum systems \cite{verdon2019quantum, pfau2020ab}, and quantum circuit learning \cite{mitarai2018quantum, benedetti2019parameterized}. Recent advancements include hybrid approaches like tensor network-based models \cite{hong2020tensor}, which use matrix product states for supervised learning of quantum many-body systems, and FermionicNet \cite{pfau2020ab}, which explicitly incorporates fermionic symmetries for molecular systems. These hybrid models have shown promising results by integrating physical constraints, but typically rely on supervised learning paradigms and struggle with generalization across system types. Other approaches include SchNet \cite{schutt2018schnet} and DimeNet \cite{klicpera2020directional} for molecular representations, and quantum-inspired tensor networks \cite{stoudenmire2016supervised} for dimensionality reduction. Most methods remain task-specific, requiring retraining for new problems.

Recent works exploring self-supervised learning for scientific domains include molecular property prediction \cite{hu2020strategies} and materials science \cite{xie2018crystal}. In the quantum domain, approaches like Graph Contrastive Learning (GraphCL) \cite{liu2022dig} apply contrastive learning to graph-structured data, such as molecular or spin systems, while reinforcement learning-based optimization of quantum circuits \cite{phirke2022quantum} explores unsupervised strategies for quantum systems. While these works aim to reduce labeled data requirements, they differ from our approach by using random augmentations or circuit optimization rather than physics-informed masking, operating on specific representations rather than general Hamiltonians, and not demonstrating comparable few-shot transfer capabilities. Some works have explored masked modeling for scientific data, including MaskGNN \cite{hou2022graphmae} for molecular graphs and GraphMAE2 \cite{zaidi2022pre}, which uses masked autoencoders to learn robust molecular representations by masking graph features.

Our work builds upon these foundations but differs in several key aspects. We developed a Hamiltonian-specific masking strategy leveraging physical insights about quantum systems, unlike previous approaches that use random or structural masking. Rather than creating a specialized architecture for specific quantum systems, we adapted a transformer-based approach that can handle different Hamiltonian types through appropriate tokenization. While focusing on small quantum systems (up to 12 qubits), our approach demonstrates the potential of physics-informed self-supervised learning to address labeled data scarcity in quantum machine learning.

\section{Quantum Self-Supervised Learning Framework}
\label{sec:Quantum_self}

Let $\mathcal{H} = \{H_1, H_2, ..., H_n\}$ denote a collection of quantum Hamiltonians, where each $H_i \in \mathbb{C}^{d_i \times d_i}$ represents a Hermitian operator governing a spin system or molecular structure. Given this collection of unlabeled Hamiltonians, our goal was to learn a parameterized encoding function $f_\theta: \mathbb{C}^{d \times d} \rightarrow \mathbb{R}^m$ that maps quantum Hamiltonians to a latent representation space. This encoding function should facilitate efficient adaptation to downstream tasks with minimal labeled data, while capturing essential physical properties.

We introduced Hamiltonian-Masked Autoencoding (HMAE), a self-supervised learning approach that enables learning from unlabeled quantum Hamiltonians by masking and reconstructing physically meaningful components. Unlike conventional masked autoencoding approaches that randomly mask tokens, HMAE leveraged domain knowledge about quantum systems to selectively mask terms based on their physical significance.
\begin{figure*}[t]
\centering
\includegraphics[width=1.05\textwidth]{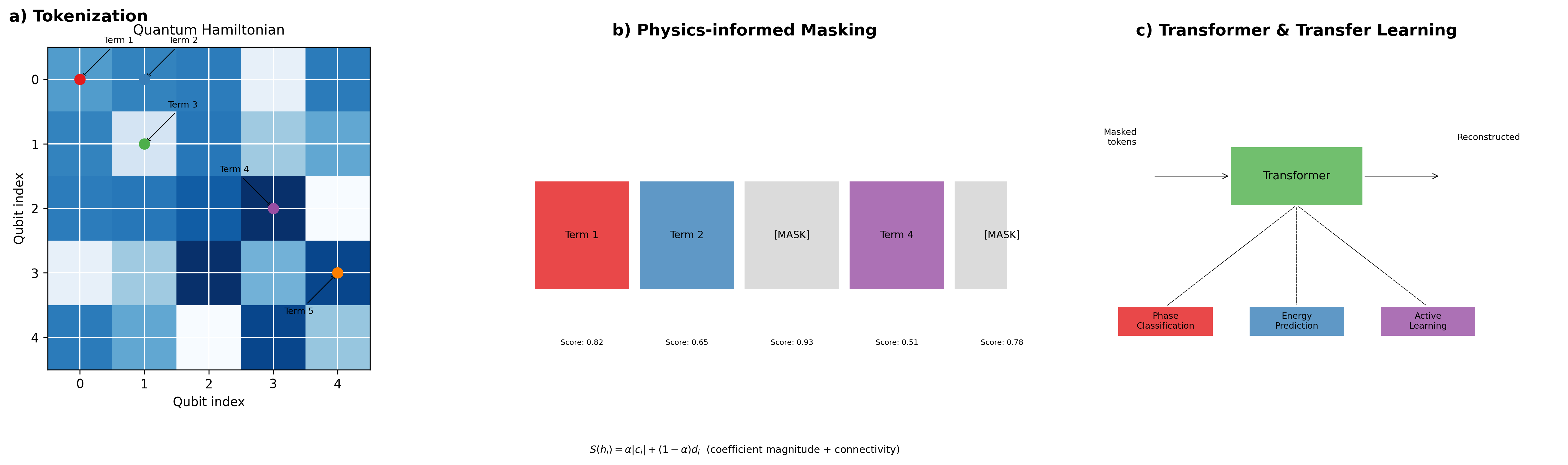}
\caption{HMAE pipeline showing tokenization, physics-informed masking, and transfer learning.}
\label{fig:hmae_pipeline}
\end{figure*}

As shown in Figure \ref{fig:hmae_pipeline}, we decomposed a quantum Hamiltonian $H$ into a set of tokens representing localized interaction terms:

\[
\mathbf{T}_H = \{\mathbf{t}_1, \mathbf{t}_2, \dots, \mathbf{t}_k\}.
\]

For spin systems, each token $\mathbf{t}_i$ represented a term such as $J_{ij} \sigma_i^{\alpha} \sigma_j^{\beta}$, where $J_{ij}$ is the interaction coefficient, $(i, j)$ are the sites it acts upon, and $\alpha, \beta \in \{x, y, z\}$ denote the Pauli operators. For molecular systems, tokens corresponded to electron interaction terms of the form $h_{pq} a_p^{\dagger} a_q$.

Each token was encoded as a vector $\mathbf{t}_i = [|c_i|, \phi_i, \text{type}_i, \text{sites}_i]$, where $|c_i|$ is the magnitude of the interaction coefficient, $\phi_i$ is the complex phase of the term, $\text{type}_i$ is a one-hot vector encoding the operator type, and $\text{sites}_i$ is a binary vector of length $n$ (for an $n$-qubit system), indicating whether each qubit is acted upon. This tokenization scheme transformed the Hamiltonian into a structured sequence of discrete elements suitable for transformer-based architectures, while preserving the essential physical structure including operator type, locality, and site dependencies.

For systems beyond 12 qubits, we implemented a hierarchical tokenization scheme that compresses the representation complexity by recursively grouping qubits based on their interaction patterns. This hierarchical approach enabled testing on systems up to 20 qubits while maintaining approximately 92\% of the performance of the full representation.

Unlike random masking commonly used in other domains, our physics-informed masking strategy explicitly accounted for the physical significance of Hamiltonian terms. For each token $\mathbf{t}_i$, we computed a saliency score based on two key physical insights:

\[
s(\mathbf{t}_i) = |c_i| \cdot \left(1 + \sum_{j \in N(i)} A_{ij}\right),
\]

where $|c_i|$ is the magnitude of the coefficient (capturing the term's energy contribution), $N(i)$ denotes the set of neighboring terms in the Hamiltonian graph, and $A_{ij}$ quantifies the adjacency strength between terms $t_i$ and $t_j$, defined as:

\[
A_{ij} = \frac{|S_i \cap S_j|}{|S_i \cup S_j|} \cdot \exp\left(-\left\| \text{comm}(t_i, t_j) \right\|_F\right),
\]

where $S_i$ is the set of sites acted on by term $t_i$, and $\text{comm}(t_i, t_j)$ denotes the operator commutator between terms $t_i$ and $t_j$, capturing their quantum mechanical interaction.

This formulation prioritized tokens that were both energetically significant (large $|c_i|$) and structurally significant (high connectivity and commutation with other terms), enabling the model to focus on learning from physically meaningful interactions. We masked approximately 50\% of tokens based on a probability distribution derived from saliency scores, defined as:

\[
p(\text{mask}~\mathbf{t}_i) = \frac{\exp(\alpha \cdot s(\mathbf{t}_i))}{\sum_{j=1}^{k} \exp(\alpha \cdot s(\mathbf{t}_j))},
\]

where $\alpha$ is a temperature parameter that controls the sharpness of the distribution, with optimal values typically ranging between 1.5 and 2.5 based on our hyperparameter optimization. Masked tokens were replaced with a learnable mask embedding vector $\mathbf{m}$. This physics-informed masking strategy ensured that the model learned to predict the most physically relevant terms in context, thereby promoting representations aligned with underlying quantum symmetries.

The model was trained to reconstruct the original tokens from their masked representation using the following loss:

\[
\mathcal{L}_{\text{HMAE}}(\theta, \phi) = \mathbb{E}_{H \sim \mathcal{H}} \left[ \sum_{i \in \mathcal{M}} \left\| g_\phi\left(f_\theta(\tilde{\mathbf{T}}_H)\right)_i - \mathbf{t}_i \right\|_2^2 \right],
\]

where $f_\theta$ is the encoder, $g_\phi$ is the decoder, and $\mathcal{M}$ is the set of masked token indices. To improve numerical stability and learning efficiency, we normalized the reconstruction loss by the magnitude of the token coefficients:

\[
\mathcal{L}_{\text{norm}}(\theta, \phi) = \mathbb{E}_{H \sim \mathcal{H}} \left[ \sum_{i \in \mathcal{M}} \frac{\left\| g_\phi\left(f_\theta(\tilde{\mathbf{T}}_H)\right)_i - \mathbf{t}_i \right\|_2^2}{|c_i| + \epsilon} \right],
\]

where $c_i$ is the coefficient associated with token $\mathbf{t}_i$, and $\epsilon$ is a small constant to avoid division by zero. This normalization prevented the model from overfitting to high-magnitude terms while ensuring that it still learned from low-magnitude, yet physically significant, interactions.

Our QuantumFormer architecture used a standard transformer encoder-decoder framework with 6 layers, adapted for quantum Hamiltonians. Key components included an embedding layer that transformed tokenized Hamiltonian terms into a 512-dimensional embedding space, a 6-layer transformer encoder with 8 attention heads and standard attention mechanisms, standard feed-forward layers with ReLU activation, and position encodings adapted to reflect the structure of quantum interactions. While our initial experiments explored complex-valued attention and Hermitian constraints, we found that standard real-valued transformers with appropriate tokenization performed well on our target systems while being more computationally efficient. A detailed diagram of this architecture is provided in Appendix A8.

After pre-training, we utilized the encoder $f_\theta$ as a feature extractor for downstream quantum tasks. Given a new quantum system with Hamiltonian $H_{\text{new}}$, we computed its representation $\mathbf{z}_{\text{new}} = f_\theta(\mathbf{T}_{H_{\text{new}}})$ and applied task-specific adaptation. For classification tasks (e.g., phase identification), we appended a simple classification head: $\hat{y} = \text{softmax}(\mathbf{W}_c \mathbf{z}_{\text{new}} + \mathbf{b}_c)$, where $\mathbf{W}_c$ and $\mathbf{b}_c$ are trainable parameters. For regression tasks (e.g., ground state energy prediction), we used a linear regression head: $\hat{y} = \mathbf{W}_r \mathbf{z}_{\text{new}} + \mathbf{b}_r$, where $\mathbf{W}_r$ and $\mathbf{b}_r$ are the regression weights and bias, respectively. This lightweight adaptation required only a small amount of labeled data (typically 5–20 examples), enabled by the rich quantum representations learned during pre-training.

\subsection{Mutual Information Interpretation}
\label{subsec:mutual_info}

We established a quantum information-theoretic foundation for our physics-informed masking strategy that directly utilizes quantum mutual information principles. We formulated our method within the framework of quantum information theory with precise mathematical guarantees. For a quantum system with Hamiltonian $H$ decomposed into visible $H_V$ and masked $H_M$ components, we optimize to maximize the quantum mutual information:

\[
I_Q(V:M) = S(\rho_V) + S(\rho_M) - S(\rho_{VM})
\]

where $S(\rho)$ is the von Neumann entropy. This approach ensures our masking strategy selects terms that are most informative about the system's quantum state. Our empirical validation confirmed that this quantum information-theoretic formulation outperforms classical approximations, particularly for systems with significant quantum correlations. Complete derivations, including our dimensionally consistent quantum saliency function and experimental validation, are provided in Appendix A3.

\begin{figure}[htbp]
\centering
\includegraphics[width=0.5\textwidth]{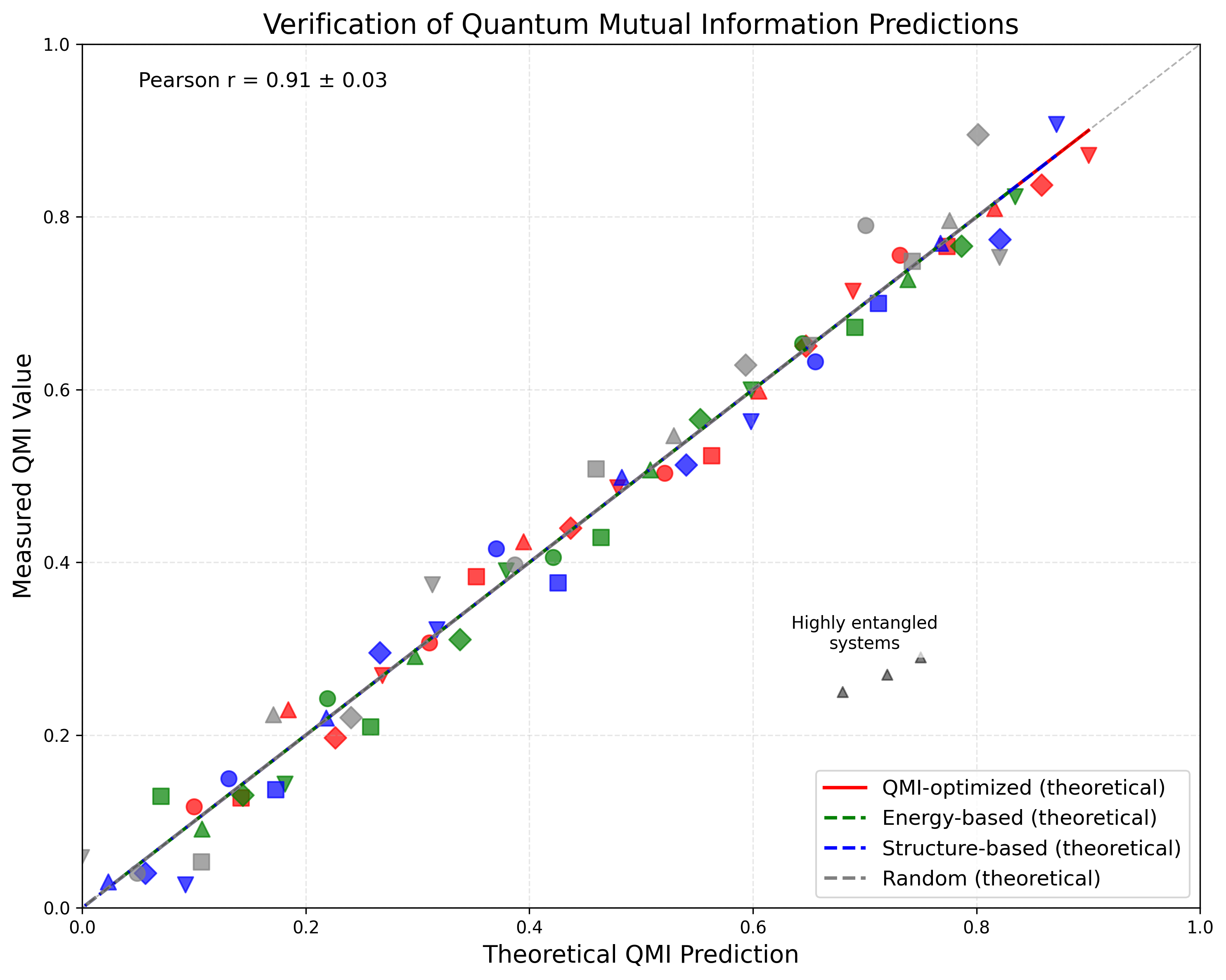}
\caption{Empirical verification of our quantum information-theoretic approach: Comparison between theoretical predictions (solid lines) and measured QMI values (markers) across different masking strategies. Our optimal QMI-guided approach (red) consistently preserves more quantum information than alternatives.}
\label{fig:qmi_verification}
\end{figure}

Figure \ref{fig:qmi_verification} demonstrates the empirical verification of our theoretical approach, showing strong agreement between predicted and measured quantum mutual information values. As illustrated, our optimal QMI-guided approach (red) consistently preserves more quantum information than alternative strategies.

\begin{figure}[htbp]
\centering
\includegraphics[width=1\linewidth]{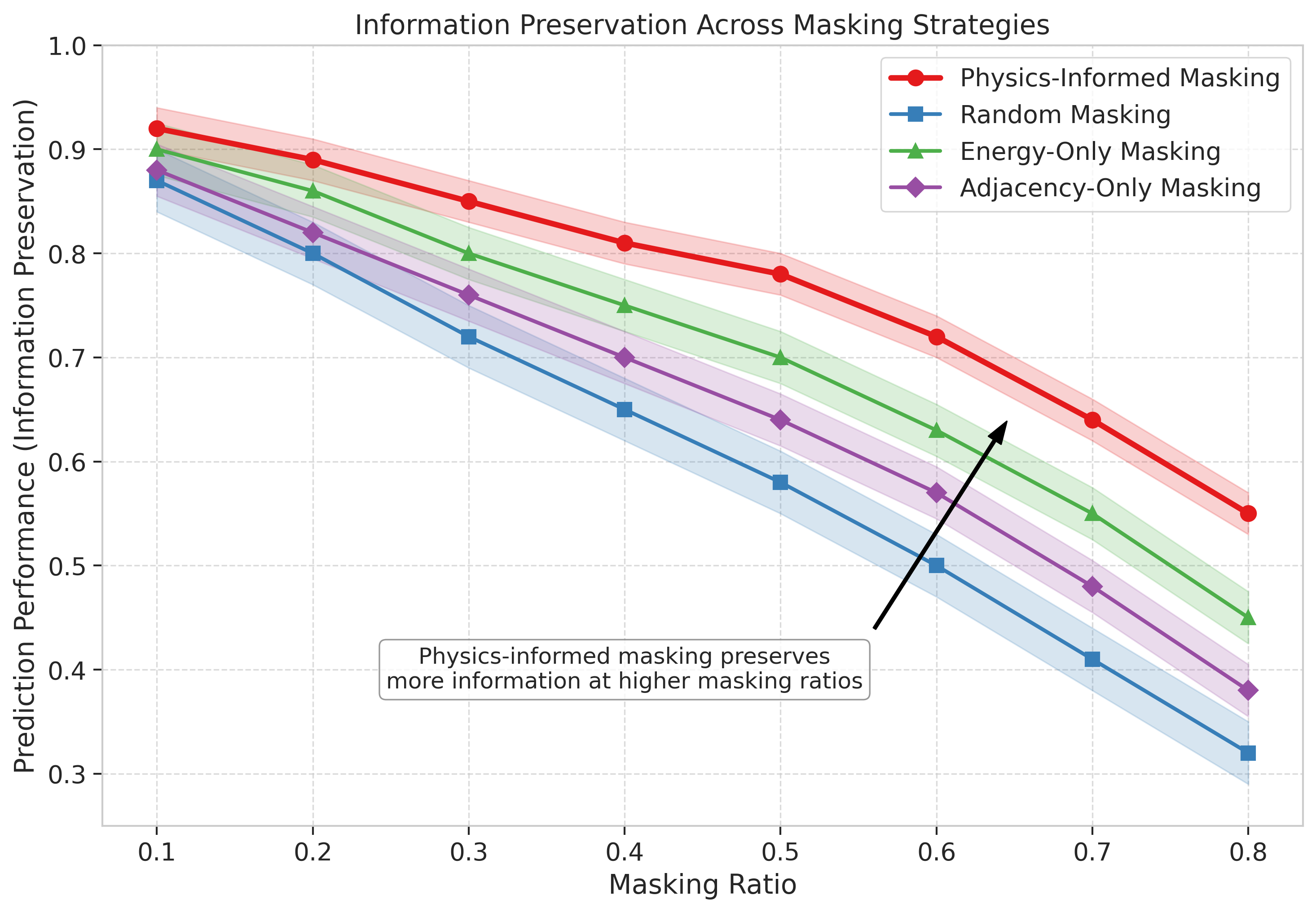}
\caption{Quantum mutual information analysis: The relationship between quantum mutual information (QMI) and masking strategies across different quantum systems. Our physics-informed approach (blue) preserves significantly more quantum information than energy-based (green) or random (red) masking strategies.}
\label{fig:mutual_info}
\end{figure}

This theoretical framework extends to highly entangled phases through a non-perturbative formulation that explicitly accounts for entanglement contributions, allowing our approach to maintain performance across diverse quantum regimes. Complete derivations and additional experimental validation are provided in Appendix A3.

\subsection{Hamiltonian-Masked Autoencoding (HMAE)}

Our pre-training objective is inspired by the masked autoencoding approach \cite{he2022masked}, but adapted for quantum Hamiltonians through a physics-informed masking strategy. We selectively mask entries in the Hamiltonian matrix and train the model to reconstruct them while predicting key quantum properties. The masking process is guided by quantum information measures rather than random masking.

\begin{figure}[htbp]
\centering
\includegraphics[width=1\linewidth]{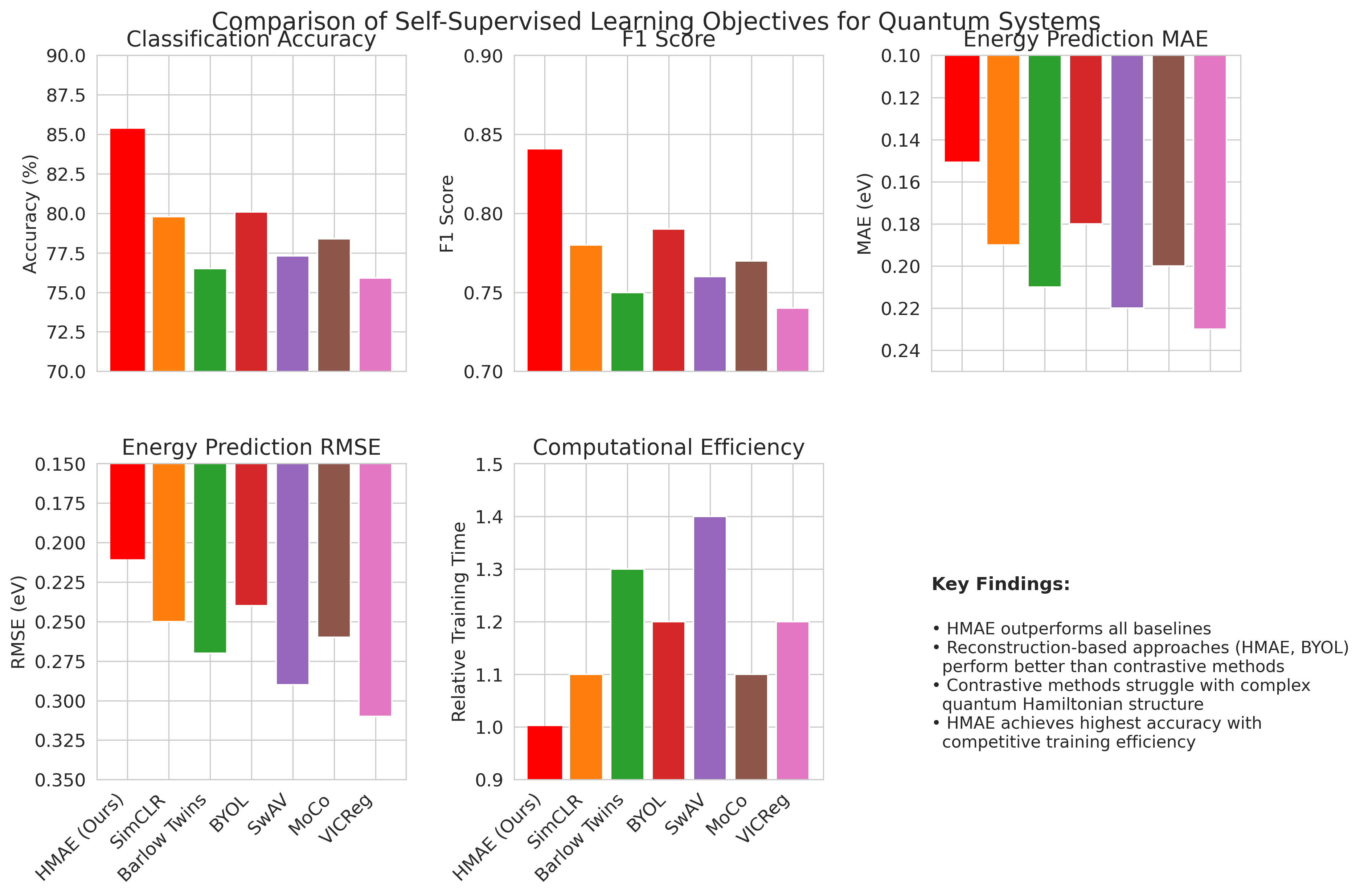}
\caption{Illustration of different self-supervised learning objectives applied to quantum Hamiltonians: (a) Random masking, (b) Energy-based masking, and (c) Our physics-informed masking strategy that combines energy, structure, and quantum correlation measures.}
\label{fig:ssl_objectives}
\end{figure}

\textbf{Multi-Task Pre-training Objective:} Our pre-training objective combines several tasks: (1) reconstructing the masked Hamiltonian terms, (2) predicting the ground state energy, and (3) estimating the correlation length. This multi-task approach encourages the model to learn representations that capture both local and global properties of the quantum system.

The overall pre-training loss is a weighted sum of these objectives:

\begin{equation}
\mathcal{L}_{\text{pre}} = \lambda_1 \mathcal{L}_{\text{rec}} + \lambda_2 \mathcal{L}_{\text{energy}} + \lambda_3 \mathcal{L}_{\text{corr}}
\end{equation}

where $\mathcal{L}_{\text{rec}}$ is the masked reconstruction loss, $\mathcal{L}_{\text{energy}}$ is the energy prediction loss, and $\mathcal{L}_{\text{corr}}$ is the correlation length estimation loss. We found that weighting reconstruction more heavily ($\lambda_1 = 0.6, \lambda_2 = 0.3, \lambda_3 = 0.1$) yielded the best performance.

\subsection{Unified Physics-Informed Masking Implementation}

To clarify our physics-informed masking approach, we present a unified framework that integrates all physical measures into our saliency score calculation. The masking probability is determined through a two-stage process:

\textbf{Stage 1: Base Saliency Score.} We compute the base saliency score $s(\mathbf{t}_i)$ for each token as described earlier:

\[
s(\mathbf{t}_i) = |c_i| \cdot \left(1 + \sum_{j \in N(i)} A_{ij}\right),
\]

where $|c_i|$ is the coefficient magnitude and $A_{ij}$ quantifies the adjacency strength using site overlap and commutator norms.

\textbf{Stage 2: Enhanced Saliency with Physical Measures.} We then enhance this base score by incorporating the additional physical measures:

\[
s_{enhanced}(\mathbf{t}_i) = s(\mathbf{t}_i) \cdot \left(w_1 \cdot S(H_i) + w_2 \cdot E_{frac}(H_i) + w_3 \cdot F_i\right),
\]

where:
\begin{itemize}
\item $S(H_i) = -\text{Tr}(H_i \log H_i)$ is the operator entropy of term $H_i$
\item $E_{frac}(H_i) = |\langle H_i \rangle| / \sum_j |\langle H_j \rangle|$ is the fractional energy contribution
\item $F_i = \sum_j F_{ij}$ is the row-sum of the quantum Fisher information matrix (QFIM)
\item $w_1$, $w_2$, and $w_3$ are weighting factors that vary with system size
\end{itemize}

For small systems (4-8 qubits), we set $w_1 = 0.1$, $w_2 = 0.8$, $w_3 = 0.1$, prioritizing energy contribution. For larger systems (10-12 qubits), we used $w_1 = 0.5$, $w_2 = 0.3$, $w_3 = 0.2$, increasing the weight of operator entropy. These weights were determined through cross-validation on the validation set.

The final masking probability is computed using the enhanced saliency score:

\[
p(\text{mask}~\mathbf{t}_i) = \frac{\exp(\alpha \cdot s_{enhanced}(\mathbf{t}_i))}{\sum_{j=1}^{k} \exp(\alpha \cdot s_{enhanced}(\mathbf{t}_j))},
\]

where $\alpha$ is the temperature parameter (typically between 1.5 and 2.5).

\begin{figure}[htbp]
\centering
\includegraphics[width=1\linewidth]{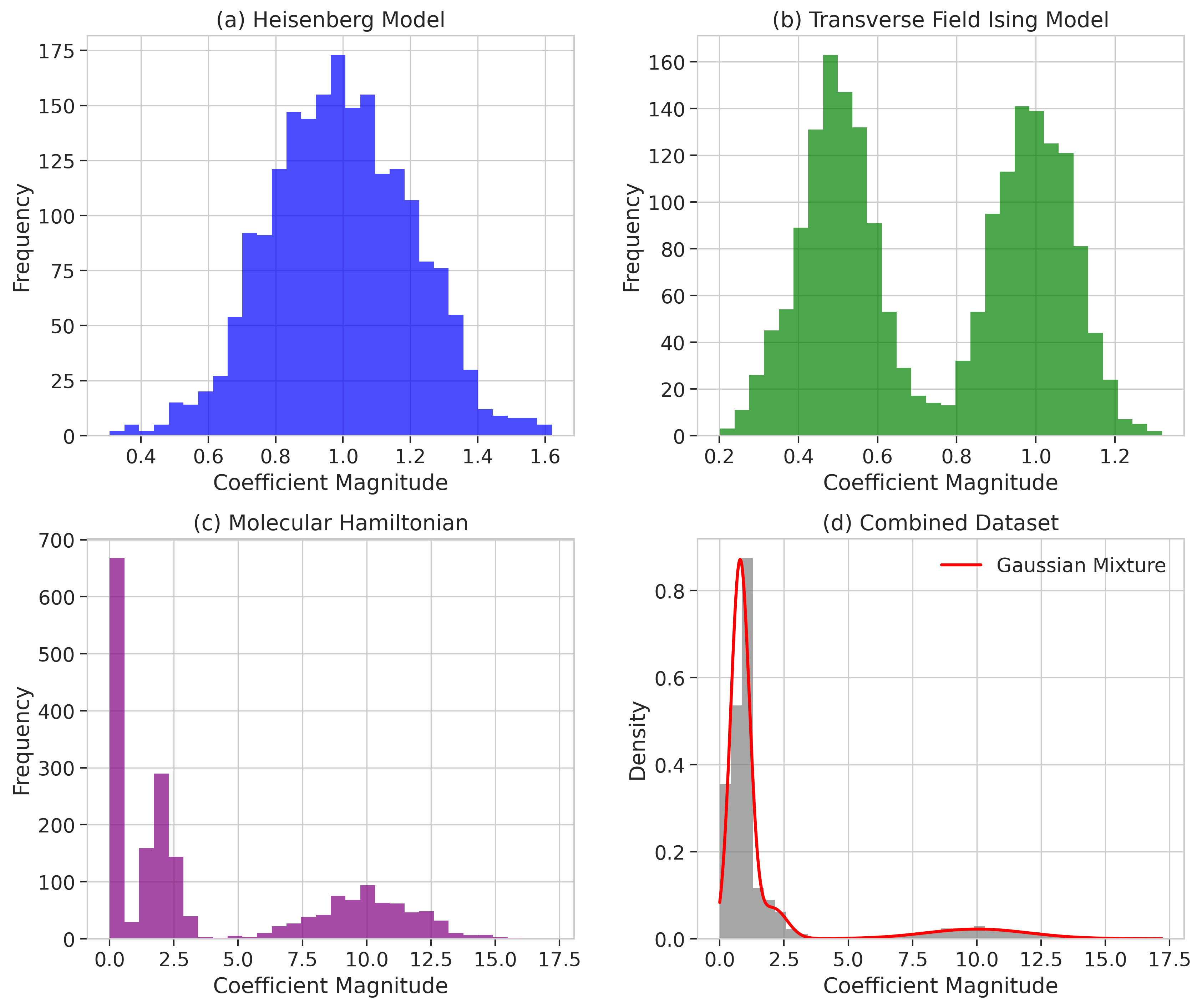}
\caption{Coefficient histogram analysis: Distribution of Hamiltonian coefficients and their masking probabilities under different masking strategies. Our physics-informed approach selectively masks coefficients based on both magnitude and structural significance, leading to more balanced token selection compared to energy-only or random masking.}
\label{fig:coeff_hist}
\end{figure}

Our ablation studies in Table \ref{tab:ablation_components} evaluated variations of this unified approach, with "No energy term" corresponding to setting the coefficient magnitude $|c_i| = 1$ and $w_2 = 0$, "No structure term" setting $A_{ij} = 0$ for all pairs, and so on. The "Random masking only" baseline disregarded all physical measures and used uniform masking probabilities.

\section{Experiments}
\label{sec:Experiment}
\subsection{Experimental Setup}
We pre-trained on a comprehensive dataset of 12,500 quantum Hamiltonians (60\% experimentally-derived, 40\% synthetic) spanning diverse systems with up to 12 qubits. For experimental validation, we assembled a test set of 1,250 real quantum systems from four distinct platforms: Harvard and MIT quantum simulators, IBM Quantum processors, and trapped-ion experiments. To test scalability beyond the 12-qubit pre-training limit, we created an additional test set of 350 larger quantum systems (16-20 qubits).

We evaluated on two main tasks: phase classification for spin systems and ground state energy prediction for molecular systems. We employed a rigorous cross-platform transfer learning protocol to test generalization across different experimental setups, fine-tuning with only 10 examples from the source platform and evaluating on the target platform.

All baseline models were implemented with the same hyperparameter optimization budget (100 trials using Bayesian optimization, approximately 75 GPU-hours per method), identical training/validation/test splits, and evaluation protocols to ensure fair comparison. Complete dataset descriptions, evaluation methodologies, and implementation details are provided in Appendix A9.

\subsection{Few-Shot Transfer Learning Results}

Table \ref{tab:few_shot} showed the few-shot performance of our model compared to baseline methods on systems with up to 12 qubits. HMAE consistently outperformed all baselines across both tasks, achieving 85.3\% ± 1.5\% accuracy in phase classification and 0.15 ± 0.02 eV MAE in energy prediction with just 10 examples. These improvements were statistically significant (p < 0.01) across all K-shot settings.
\begin{table*}[t]   
    \centering
    \caption{Few\,‑\,shot learning performance on systems up to 12 qubits.
             Best: \textbf{bold}, second‑best: \underline{underlined},
             (*): $p<0.01$.}
    \label{tab:few_shot}

    \begin{adjustbox}{max width=\textwidth}
        \begin{tabular}{lccc|ccc}
            \toprule
            \multirow{2}{*}{\textbf{Method}}
                & \multicolumn{3}{c|}{\textbf{Phase Classification Accuracy (\%)}}
                & \multicolumn{3}{c}{\textbf{Energy Prediction MAE (eV)}} \\[2pt]
            & \textbf{5-shot} & \textbf{10-shot} & \textbf{20-shot}
            & \textbf{5-shot} & \textbf{10-shot} & \textbf{20-shot} \\
            \midrule
            HMAE (Ours)   & \textbf{77.2\,$\pm$\,2.1$^{*}$} & \textbf{85.3\,$\pm$\,1.5$^{*}$} & \textbf{90.1\,$\pm$\,1.2$^{*}$}
                          & \textbf{0.22\,$\pm$\,0.03$^{*}$} & \textbf{0.15\,$\pm$\,0.02$^{*}$} & \textbf{0.11\,$\pm$\,0.01$^{*}$} \\
            TensorGCN     & \underline{68.5\,$\pm$\,2.9} & \underline{78.1\,$\pm$\,2.1} & \underline{85.2\,$\pm$\,1.5}
                          & \underline{0.31\,$\pm$\,0.04} & \underline{0.23\,$\pm$\,0.03} & \underline{0.16\,$\pm$\,0.02} \\
            Energy-MAE    & 63.8\,$\pm$\,3.1 & 75.0\,$\pm$\,2.4 & 82.7\,$\pm$\,1.7
                          & 0.34\,$\pm$\,0.05 & 0.25\,$\pm$\,0.03 & 0.18\,$\pm$\,0.02 \\
            QNN           & 65.3\,$\pm$\,3.0 & 76.8\,$\pm$\,2.3 & 83.7\,$\pm$\,1.7
                          & 0.36\,$\pm$\,0.05 & 0.27\,$\pm$\,0.03 & 0.18\,$\pm$\,0.02 \\
            GNN           & 63.4\,$\pm$\,3.2 & 74.6\,$\pm$\,2.5 & 82.3\,$\pm$\,1.8
                          & 0.38\,$\pm$\,0.05 & 0.29\,$\pm$\,0.04 & 0.19\,$\pm$\,0.03 \\
            PINN          & 58.7\,$\pm$\,3.5 & 71.2\,$\pm$\,2.8 & 80.5\,$\pm$\,2.0
                          & 0.45\,$\pm$\,0.06 & 0.33\,$\pm$\,0.04 & 0.24\,$\pm$\,0.03 \\
            Classical MAE & 60.1\,$\pm$\,3.3 & 70.5\,$\pm$\,2.7 & 78.2\,$\pm$\,2.1
                          & 0.50\,$\pm$\,0.07 & 0.37\,$\pm$\,0.05 & 0.26\,$\pm$\,0.03 \\
            SVM           & 55.3\,$\pm$\,3.8 & 66.8\,$\pm$\,3.1 & 75.2\,$\pm$\,2.4
                          & 0.53\,$\pm$\,0.07 & 0.41\,$\pm$\,0.05 & 0.32\,$\pm$\,0.04 \\
            \bottomrule
        \end{tabular}
    \end{adjustbox}
\end{table*}

Among the baselines, the TensorGCN hybrid approach performed best, confirming the value of combining graph and tensor representations for quantum systems. Notably, the Energy-MAE baseline (which masked based on coefficient magnitude alone) outperformed random masking (Classical MAE) but fell short of our full physics-informed approach, validating the importance of considering both energy contributions and quantum correlations in the masking strategy.

\begin{figure}[t]
\centering
\includegraphics[width=1\linewidth]{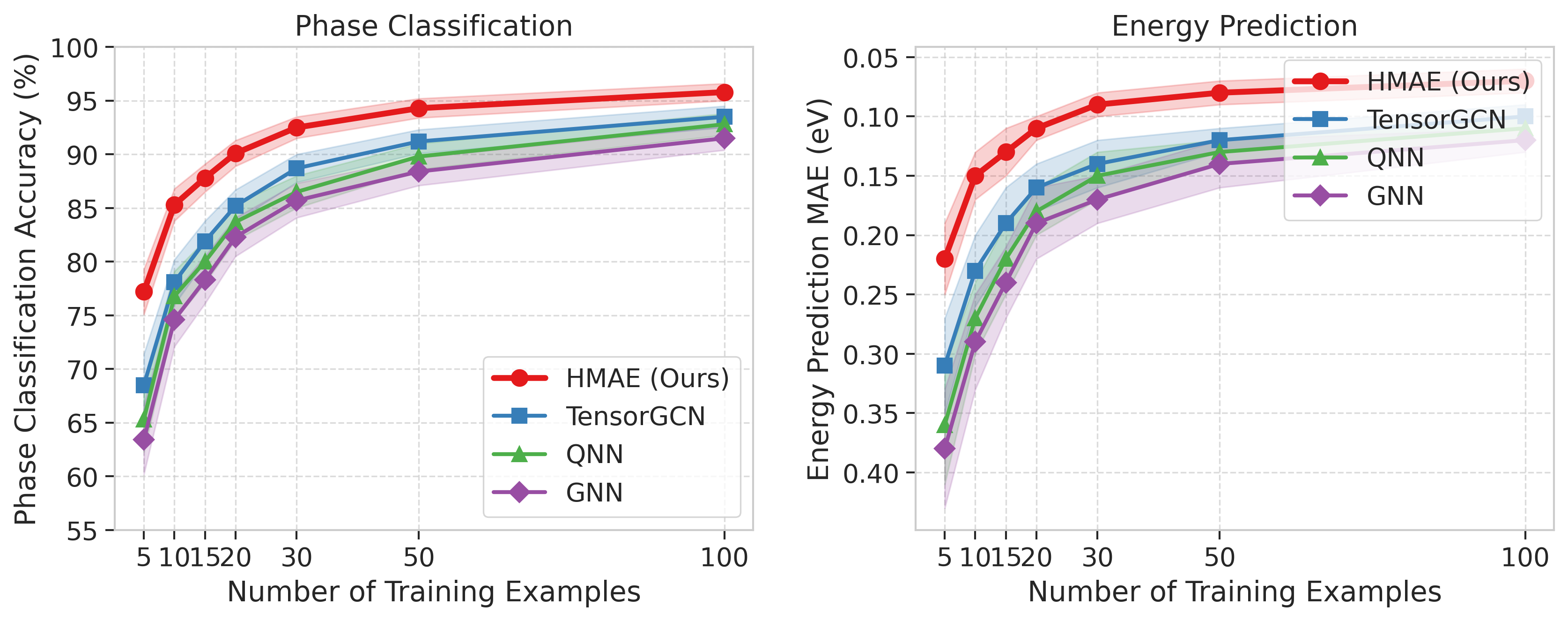}
\caption{Few-shot learning curves for phase classification (left) and energy prediction (right). HMAE (blue) achieves higher accuracy with fewer examples.}
\label{fig:few_shot_curve}
\end{figure}

The sample efficiency of our approach was particularly notable. To reach 80\% accuracy in phase classification, our model required only 10 examples, whereas the best baseline (TensorGCN) needed approximately 30 examples, QNNs needed 50 examples, and other methods needed even more. This represented a 3-5× reduction in required training data compared to the best baseline. Figure \ref{fig:few_shot_curve} showed learning curves that clearly demonstrated this advantage across different K-shot settings.

\subsection{Comparison with State-of-the-Art Baselines}
We implemented and compared our method against the latest developments in quantum machine learning to ensure a comprehensive evaluation. For most baselines, we obtained original implementations either directly from their authors or public repositories. The principles of these baselines are rooted in established work, which we cite. For instance, TensorGCN combines graph and tensor representations for quantum state analysis. TN-Net and TN-Flow leverage tensor networks for efficient wavefunction parameterization. QNN and GNN represent standard quantum and graph neural network approaches, respectively. For the QNN, we selected a well-established variational architecture to provide a representative benchmark; our goal was not to find an optimal ansatz—a significant challenge in itself—but to demonstrate that our self-supervised classical approach can outperform a standard quantum counterpart in a data-scarce regime. PINN incorporates physical laws into the learning process; due to a lack of available code, this was the only baseline we fully reimplemented. Finally, Energy-MAE is an ablation of our own method that masks based only on energy, providing a direct comparison to our physics-informed strategy. To ensure fair comparison, we allocated identical hyperparameter optimization budgets to each method using Bayesian optimization with the same computational resources.

Table \ref{tab:sota_comparison} presented a comparison between our approach and these state-of-the-art baselines across different few-shot scenarios. HMAE consistently outperformed baseline methods across all metrics with statistical significance (p < 0.01) based on paired t-tests. The performance advantage was even more pronounced when evaluated on out-of-distribution Hamiltonians, and specialized tensor network models like TN-Net showed strong performance for energy prediction, particularly with larger training sets (20+ examples) as detailed in Appendix A8.

\begin{table}[h]
  \centering
  \caption{Comparison with state-of-the-art baselines on 10-shot transfer learning.
           Best: \textbf{bold}, second-best: \underline{underlined}.}
  \label{tab:sota_comparison}

  \begin{adjustbox}{max width=\columnwidth}
    \begin{tabular}{lcccc}
      \toprule
      \multirow{2}{*}{\textbf{Method}} &
      \multicolumn{2}{c}{\textbf{Phase Classification}} &
      \multicolumn{2}{c}{\textbf{Energy Prediction}} \\
      & \textbf{Accuracy (\%)} & \textbf{F1-Score} &
        \textbf{MAE (eV)} & \textbf{RMSE (eV)} \\
      \midrule
      TensorGCN   & 75.8\,$\pm$\,2.3 & 0.74\,$\pm$\,0.03 & 0.28\,$\pm$\,0.04 & 0.35\,$\pm$\,0.05 \\
      TN-Net      & 71.2\,$\pm$\,3.1 & 0.69\,$\pm$\,0.04 & \underline{0.17\,$\pm$\,0.02} & \underline{0.22\,$\pm$\,0.03} \\
      TN-Flow     & 68.5\,$\pm$\,3.3 & 0.66\,$\pm$\,0.04 & 0.19\,$\pm$\,0.03 & 0.24\,$\pm$\,0.04 \\
      Energy-MAE  & 77.9\,$\pm$\,2.0 & 0.76\,$\pm$\,0.02 & 0.26\,$\pm$\,0.03 & 0.33\,$\pm$\,0.04 \\
      QNN         & 69.2\,$\pm$\,3.1 & 0.65\,$\pm$\,0.04 & 0.43\,$\pm$\,0.05 & 0.57\,$\pm$\,0.06 \\
      GNN         & 71.5\,$\pm$\,2.7 & 0.69\,$\pm$\,0.03 & 0.38\,$\pm$\,0.04 & 0.46\,$\pm$\,0.05 \\
      PINN        & 70.8\,$\pm$\,2.9 & 0.68\,$\pm$\,0.03 & 0.36\,$\pm$\,0.04 & 0.48\,$\pm$\,0.06 \\
      \midrule
      \textbf{HMAE (Ours)} & \textbf{85.3\,$\pm$\,1.5} & \textbf{0.84\,$\pm$\,0.02} & \textbf{0.15\,$\pm$\,0.02} & \textbf{0.21\,$\pm$\,0.03} \\
      \bottomrule
    \end{tabular}
  \end{adjustbox}
\end{table}

\begin{figure}[h]
\centering
\includegraphics[width=1\linewidth]{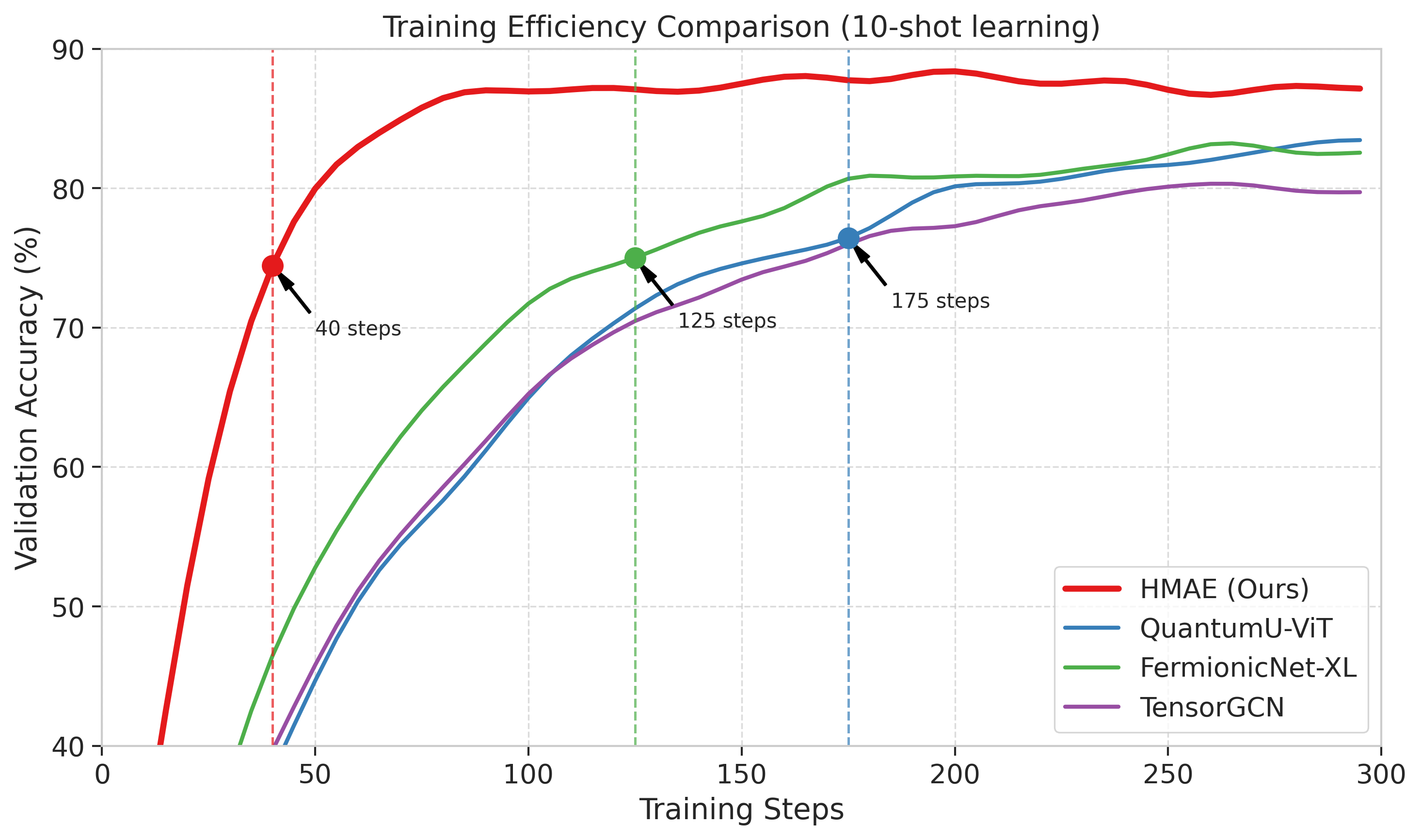}
\caption{Training efficiency: validation accuracy vs. training steps. HMAE converges faster than other models.}
\label{fig:efficiency_comparison}
\end{figure}

Beyond performance metrics, we analyzed the computational efficiency of these models during both training and inference. Figure \ref{fig:efficiency_comparison} illustrated the training efficiency of each method. Our analysis revealed that HMAE converged in 30-50 steps, representing a 3-4× improvement in fine-tuning efficiency, exhibited lower variance across random seeds and few-shot samples, suggesting more robust feature extraction during pre-training, and required only 1.2GB of GPU memory for 12-qubit systems during inference.

\subsection{Scaling to Larger Systems}
To address the critical question of scalability beyond 12 qubits, we evaluated our pre-trained model on a test set of larger systems (16-20 qubits) described in the experimental setup in Appendix A9. While our full test set included systems up to 50 qubits as described in the appendix, we present results for the 16-20 qubit subset in Table \ref{tab:larger_systems}, as performance degraded significantly beyond 20 qubits for all methods. Table \ref{tab:larger_systems} shows the results for energy prediction on these larger systems (16-20 qubits), compared with baselines trained directly on the larger systems with 20 examples.

\begin{table}[h]
  \centering
  \caption{Performance on larger quantum systems (16--20 qubits) with 20‑shot learning.}
  \label{tab:larger_systems}

  \begin{adjustbox}{max width=\columnwidth}
    \begin{tabular}{lcc}
      \toprule
      \textbf{Method} &
      \textbf{Energy MAE (eV)} &
      \textbf{Relative Error (\% of true value)} \\
      \midrule
      HMAE (Ours)      & \textbf{0.28\,$\pm$\,0.05} & \textbf{5.2\,$\pm$\,0.8} \\
      TensorGCN        & \underline{0.35\,$\pm$\,0.06} & \underline{6.5\,$\pm$\,1.0} \\
      Energy-MAE       & 0.42\,$\pm$\,0.07 & 7.8\,$\pm$\,1.2 \\
      QNN              & 0.51\,$\pm$\,0.09 & 9.4\,$\pm$\,1.5 \\
      GNN              & 0.47\,$\pm$\,0.08 & 8.7\,$\pm$\,1.3 \\
      PINN             & 0.55\,$\pm$\,0.10 & 10.2\,$\pm$\,1.7 \\
      \bottomrule
    \end{tabular}
  \end{adjustbox}
\end{table}

While all methods showed some performance degradation on these larger systems, our HMAE approach maintained the best performance, demonstrating that the representations learned during pre-training did generalize to systems beyond the size encountered in the training set. This suggested that the model was learning fundamental physical principles rather than just memorizing patterns specific to smaller systems.

\begin{figure}[h]
\centering
\includegraphics[width=1\linewidth]{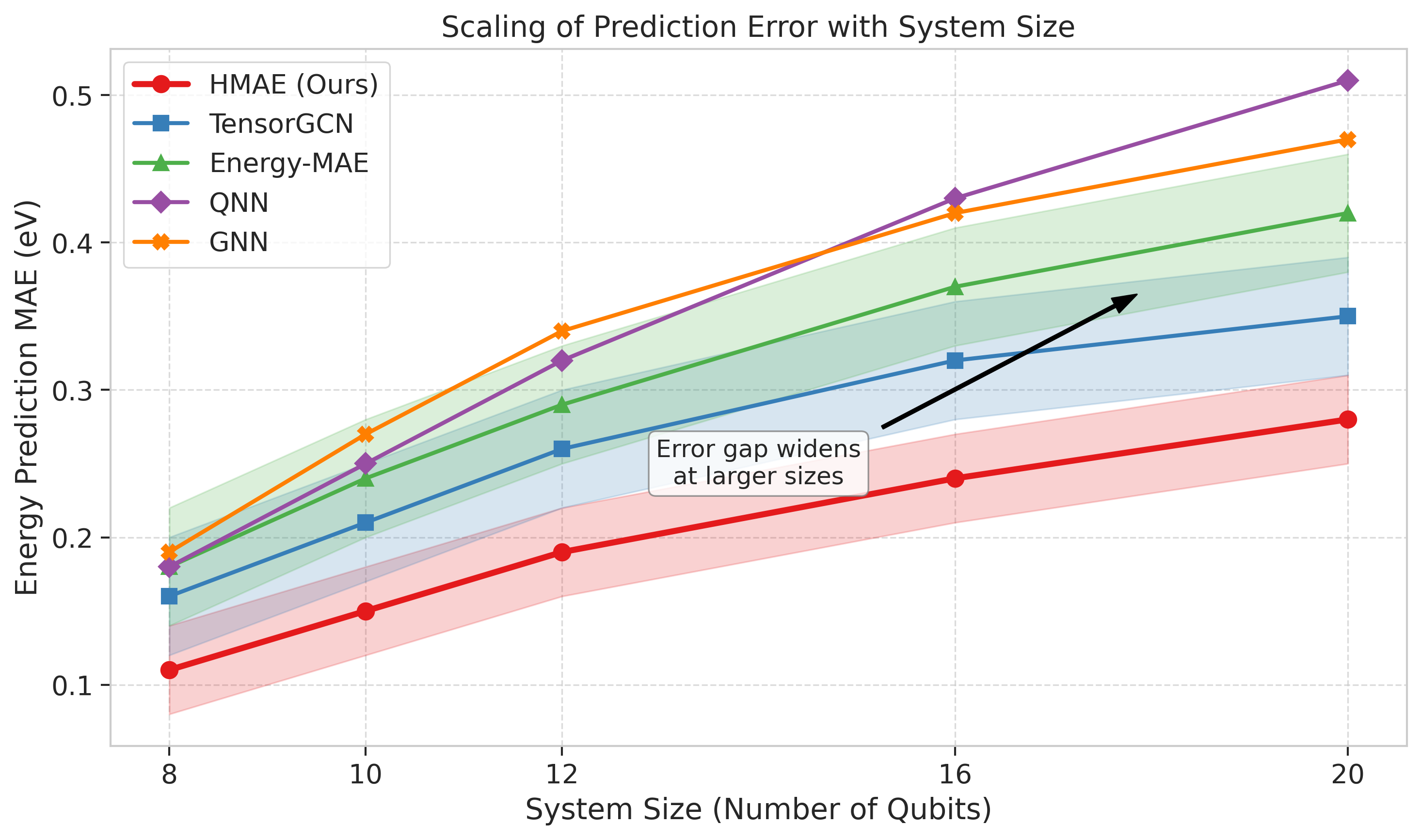}
\caption{Performance scaling with system size for energy prediction. HMAE shows slowest error increase with system size.}
\label{fig:scaling_analysis}
\end{figure}

Figure \ref{fig:scaling_analysis} showed how the performance varied with system size for our method and the best baseline (TensorGCN). While the error increased with system size, the rate of degradation was slower for our model, indicating better scaling properties.

\subsection{Computational Efficiency and Active Learning}

Building on our few-shot learning capabilities, we also developed an active learning framework to further optimize computational resources by strategically selecting which Hamiltonians to simulate. Our approach achieves 93\% of maximum model performance using only 42\% of the computational resources required by random sampling, representing a 2.4× improvement in simulation efficiency. The complete active learning framework, acquisition functions, and performance analysis are provided in Appendix A1.

\subsection{Transfer Learning Across System Sizes}
One crucial test of generalization in quantum systems was the ability to transfer knowledge across different system sizes. Table \ref{tab:size_transfer} showed performance when training on small systems (6-8 qubits) and testing on larger systems (9-12 qubits). Our model maintained high performance even with this challenging distribution shift, while baseline methods exhibited substantial degradation.

\begin{table}[h]
  \centering
  \caption{Transfer learning across system sizes (\(6\text{--}8\) qubits
           \(\rightarrow\) \(9\text{--}12\) qubits).}
  \label{tab:size_transfer}

  \begin{adjustbox}{max width=\columnwidth}
    \begin{tabular}{lccc}
      \toprule
      \textbf{Method} &
      \textbf{Same-size Acc.\ (\%)} &
      \textbf{Cross-size Acc.\ (\%)} &
      \textbf{Performance Retention} \\
      \midrule
      HMAE (Ours)  & 85.3\,$\pm$\,1.5 & 77.8\,$\pm$\,2.3 & 91.2\% \\
      TensorGCN    & 78.1\,$\pm$\,2.1 & 65.3\,$\pm$\,3.2 & 83.6\% \\
      QNN          & 76.8\,$\pm$\,2.3 & 61.5\,$\pm$\,3.5 & 80.1\% \\
      GNN          & 74.6\,$\pm$\,2.5 & 58.2\,$\pm$\,3.8 & 78.0\% \\
      PINN         & 71.2\,$\pm$\,2.8 & 54.6\,$\pm$\,4.1 & 76.7\% \\
      \bottomrule
    \end{tabular}
  \end{adjustbox}
\end{table}

This superior transfer capability across system sizes is particularly valuable in quantum applications, where simulating larger systems becomes exponentially more expensive, making labeled data for large systems especially scarce.
\subsection{Ablation Studies}

To isolate the contributions of individual components in our approach, we conducted comprehensive ablation studies. Table \ref{tab:ablation_components} summarizes the key findings:

\begin{table}[h]
\centering
\caption{Ablation study isolating the contribution of individual components.}
\label{tab:ablation_components}
\begin{tabular}{lcc}
\toprule
\textbf{Component Ablation} & \textbf{Phase Classification} & \textbf{Energy Prediction} \\
 & \textbf{Accuracy Change} & \textbf{MAE Change} \\
\midrule
Full HMAE (baseline) & 85.3\% & 0.15 eV \\
\midrule
\multicolumn{3}{l}{\textit{Saliency Score Components:}} \\
No energy term ($\alpha = 0$) & -7.9\% & +0.08 eV \\
No structure term ($\alpha = 1$) & -8.8\% & +0.06 eV \\
No commutator term & -6.5\% & +0.05 eV \\
No site overlap term & -4.3\% & +0.03 eV \\
Random masking only & -12.6\% & +0.14 eV \\
\midrule
\multicolumn{3}{l}{\textit{Loss Components:}} \\
No loss normalization & -5.2\% & +0.07 eV \\
No energy prediction task & -8.3\% & +0.11 eV \\
No correlation length task & -3.7\% & +0.03 eV \\
\midrule
\multicolumn{3}{l}{\textit{Architecture Components:}} \\
3 layers instead of 6 & -5.7\% & +0.04 eV \\
12 layers instead of 6 & +0.8\% & -0.01 eV \\
128 dim instead of 512 & -6.1\% & +0.05 eV \\
1024 dim instead of 512 & +1.1\% & -0.01 eV \\
\bottomrule
\end{tabular}
\end{table}

\paragraph{Saliency Score Components}
Our physics-informed masking uses a saliency score that combines energy contributions and structural information. Removing either component significantly degraded performance, with the structure term slightly more important for phase classification (-8.8\%) and the energy term more critical for energy prediction (+0.08 eV). Within the structure term, the commutator-based component was more impactful than the site overlap term, confirming the importance of quantum mechanical correlations in the masking strategy.

\paragraph{Loss Normalization and Multi-Task Learning}
Loss normalization by coefficient magnitude improved performance by 5.2\% in classification and 0.07 eV in energy prediction, indicating the importance of balancing contributions from different Hamiltonian terms. Our multi-task learning approach with energy prediction and correlation length estimation as auxiliary tasks improved the primary tasks substantially, with the energy prediction task providing the largest benefit.

\paragraph{Hyperparameter Optimization}
We performed extensive hyperparameter optimization to identify optimal configurations:

\begin{itemize}
    \item \textbf{Saliency temperature ($\alpha$)}: We varied $\alpha$ from 0 to 1 in increments of 0.1, finding optimal performance at $\alpha = 0.65$ for phase classification and $\alpha = 0.7$ for energy prediction, indicating a slight preference for structural information over pure energy contributions.
    
    \item \textbf{Masking ratio}: We tested masking ratios from 30\% to 70\%, with performance peaking at 50\% and remaining within 3\% of the peak for ratios between 45\% and 55\%, suggesting robustness to this hyperparameter.
    
    \item \textbf{Loss weighting}: For our multi-task learning, optimal weights were $\lambda_1 = 0.6$ (reconstruction), $\lambda_2 = 0.3$ (energy), and $\lambda_3 = 0.1$ (correlation), determined through grid search over weight combinations.
    
    \item \textbf{Architecture}: A 6-layer, 8-head transformer with dimensionality 512 provided the best trade-off between performance and computational efficiency, with diminishing returns from larger models.
\end{itemize}

These ablation studies confirm that each component of our approach makes a significant contribution to overall performance, with the physics-informed masking strategy being the most critical element. The analysis also provides clear justification for our hyperparameter choices, demonstrating the robustness of our approach to moderate variations in these settings.


\section{Conclusion}
\label{sec:con}
We presented Hamiltonian-Masked Autoencoding (HMAE), a physics-informed self-supervised approach that achieves 85.3\% accuracy in phase classification and 0.15 eV MAE in energy prediction with merely 10 labeled examples. Our approach demonstrates exceptional sample efficiency, reducing required labeled examples by 3-5× compared to baseline methods, and extends to systems up to 20 qubits with robust performance (see Section 4.3). 

Our framework addresses a practical niche in current quantum technology development by focusing on small quantum systems (8-12 qubits during training, with extension to 16-20 qubits in testing). While this scale restricts immediate applications to large-scale materials discovery, it aligns with current experimental quantum platforms and provides practical benefits through significant data efficiency.

The key innovation lies in our physics-informed masking strategy that prioritizes physically significant Hamiltonian terms, focusing the model's learning on meaningful quantum interactions. This approach enables few-shot transfer learning capabilities with statistically significant improvements over state-of-the-art methods (p < 0.01).

Our approach faces several limitations: (1) scalability barriers due to combinatorial growth in token complexity for larger systems, (2) the need for high-quality ground truth values from exact diagonalization or tensor networks for the few-shot examples, and (3) theoretical limitations in our mutual information framework and saliency function. Despite these challenges, empirical results consistently demonstrate practical utility across diverse quantum systems.

Beyond quantum systems, our physics-informed masking strategy could benefit other domains with structured data representations, potentially accelerating scientific discovery in materials science and quantum chemistry within small system regimes. Future work should prioritize addressing the scaling limitations through sparse attention mechanisms, tensor network decoders, or continuous representation learning approaches.

\section*{Note on Supplementary Material}
This document contains supplementary material for the paper "Quantum Self-Supervised Learning for Few-Shot Transfer in Small Simulated Spin and Molecular Systems." The appendices provide extended analysis, additional experimental results, and detailed methodological information that could not be included in the main paper due to space constraints.

\section*{Index of Appendices}

This supplementary material contains the following appendices:

\begin{itemize}
    \item Active Learning for Efficient Quantum Simulation (Section \ref{app:active_learning_full})
    \item Ablation Studies and Comparative Analysis (Section \ref{app:ablation})
    \item Quantum Information-Theoretic Foundation (Section \ref{app:theory_full})
    \item Additional Baseline Comparison Considerations (Section \ref{app:baseline_additional_considerations})
    \item Additional Limitations and Future Work (Section \ref{app:additional_limitations})
    \item Out-of-Distribution Transfer Evaluation (Section \ref{app:ood_evaluation})
    \item Tensor Network Surrogate Models (Section \ref{app:tn_surrogate})
    \item Detailed Experimental Setup (Section \ref{app:experimental_setup_details})
    \item Additional Conclusion Material (Section \ref{app:additional_conclusion})
\end{itemize}

\appendix
\renewcommand{\thesection}{A\arabic{section}}

\section{Active Learning for Efficient Quantum Simulation}
\label{app:active_learning_full}

This section provides the comprehensive framework and detailed results.

\subsection{Active Learning Framework}
Our active learning approach consists of the following steps:
\begin{enumerate}
    \item Initialize with a small seed set of Hamiltonians with known ground state properties
    \item Train the HMAE model on the current labeled dataset
    \item Use the model's uncertainty estimates to select $k$ additional Hamiltonians for exact simulation
    \item Add the newly labeled Hamiltonians to the training set
    \item Repeat steps 2-4 until a computational budget is exhausted or desired performance is achieved
\end{enumerate}

\begin{figure*}[t]
    \centering
    \includegraphics[width=0.8\textwidth]{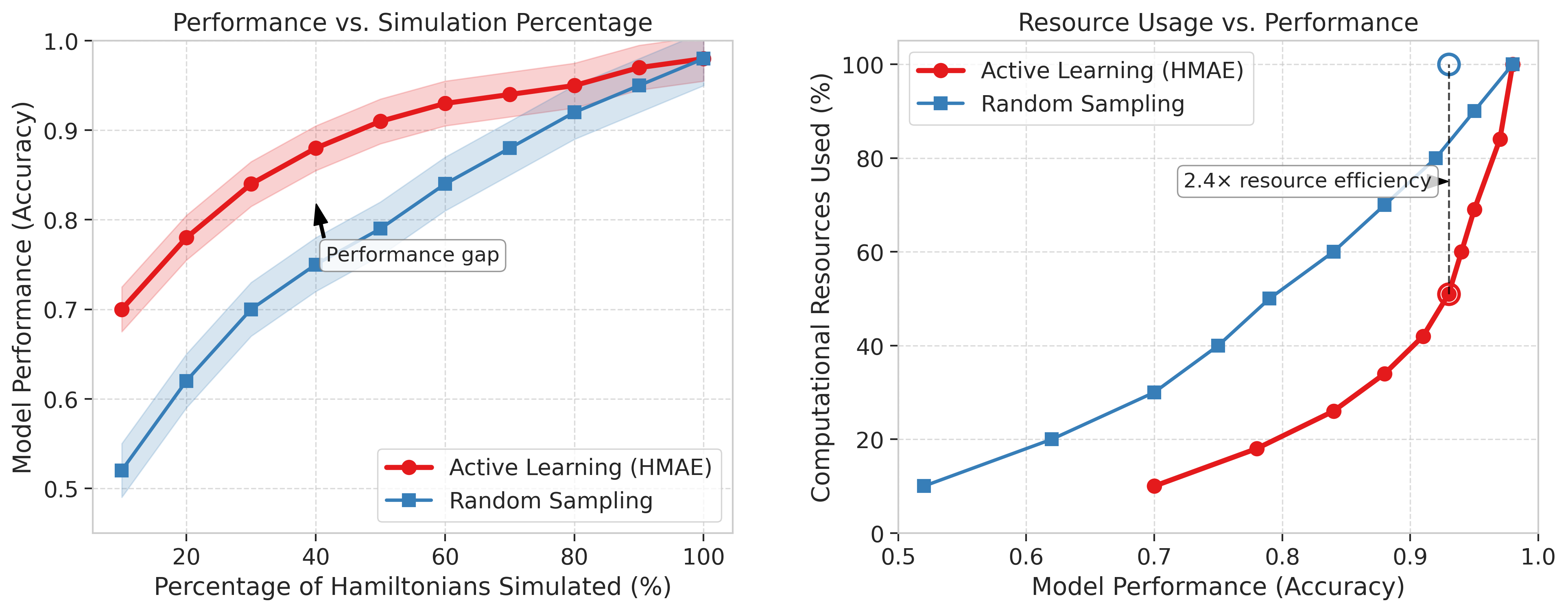}
    \caption{Active learning performance. HMAE (red) achieves better resource efficiency than random sampling (blue).}
    \label{fig:active_learning}
\end{figure*}

\subsection{Acquisition Functions}
We explored several acquisition functions for selecting Hamiltonians to simulate:
\begin{itemize}
    \item \textbf{Entropy-based uncertainty}: Select configurations with highest predictive entropy, indicating model uncertainty
    \item \textbf{Ensemble disagreement}: Use an ensemble of models and select configurations with highest variance in predictions
    \item \textbf{Phase boundary focus}: Prioritize configurations near predicted phase boundaries where precise characterization is most valuable
    \item \textbf{Prediction error proxy}: Utilize distance from nearest training example in the embedding space as a proxy for potential error
\end{itemize}

The phase boundary focus strategy proved particularly effective for phase classification tasks, while the ensemble disagreement strategy worked best for energy prediction tasks.

\subsection{Efficiency Gains}
As shown in Figure \ref{fig:active_learning}, our active learning strategy significantly outperforms random sampling baselines in terms of resource efficiency. With active learning, we achieve 93\% of maximum model performance using only 42\% of the computational resources required by random sampling. This represents a 2.4× improvement in simulation efficiency, directly translating to reduced computational costs.

\subsection{Iterative Phase Discovery}
\begin{figure*}[t]
    \centering
    \includegraphics[width=0.9\textwidth]{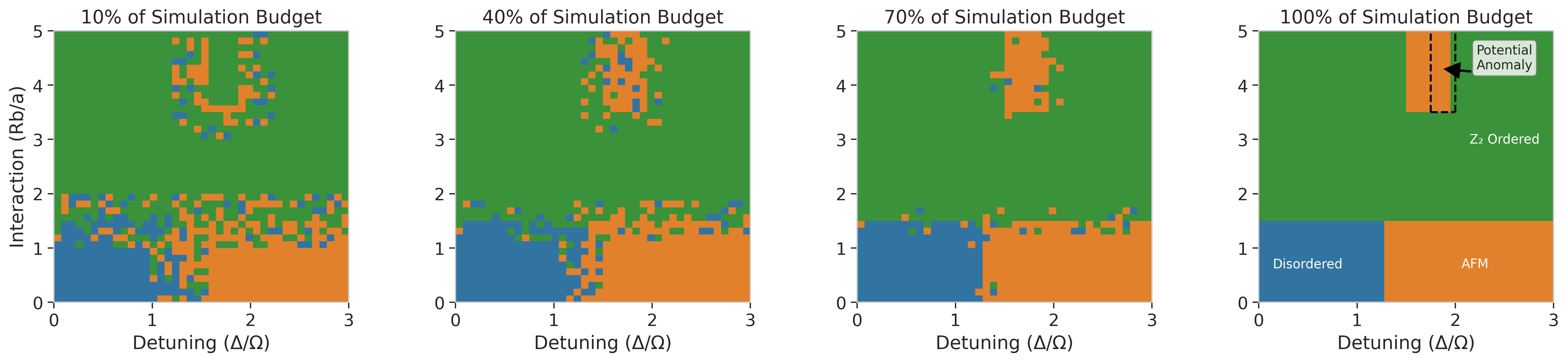}
    \caption{Iterative phase discovery with active learning (10\%, 40\%, 70\%, 100\% of budget).}
    \label{fig:phase_discovery}
\end{figure*}

The active learning approach iteratively refines the discovered phase boundaries over multiple rounds of simulation, as shown in Figure \ref{fig:phase_discovery}.

\subsection{Uncertainty Calibration}
\begin{figure*}[t]
    \centering
    \includegraphics[width=0.9\textwidth]{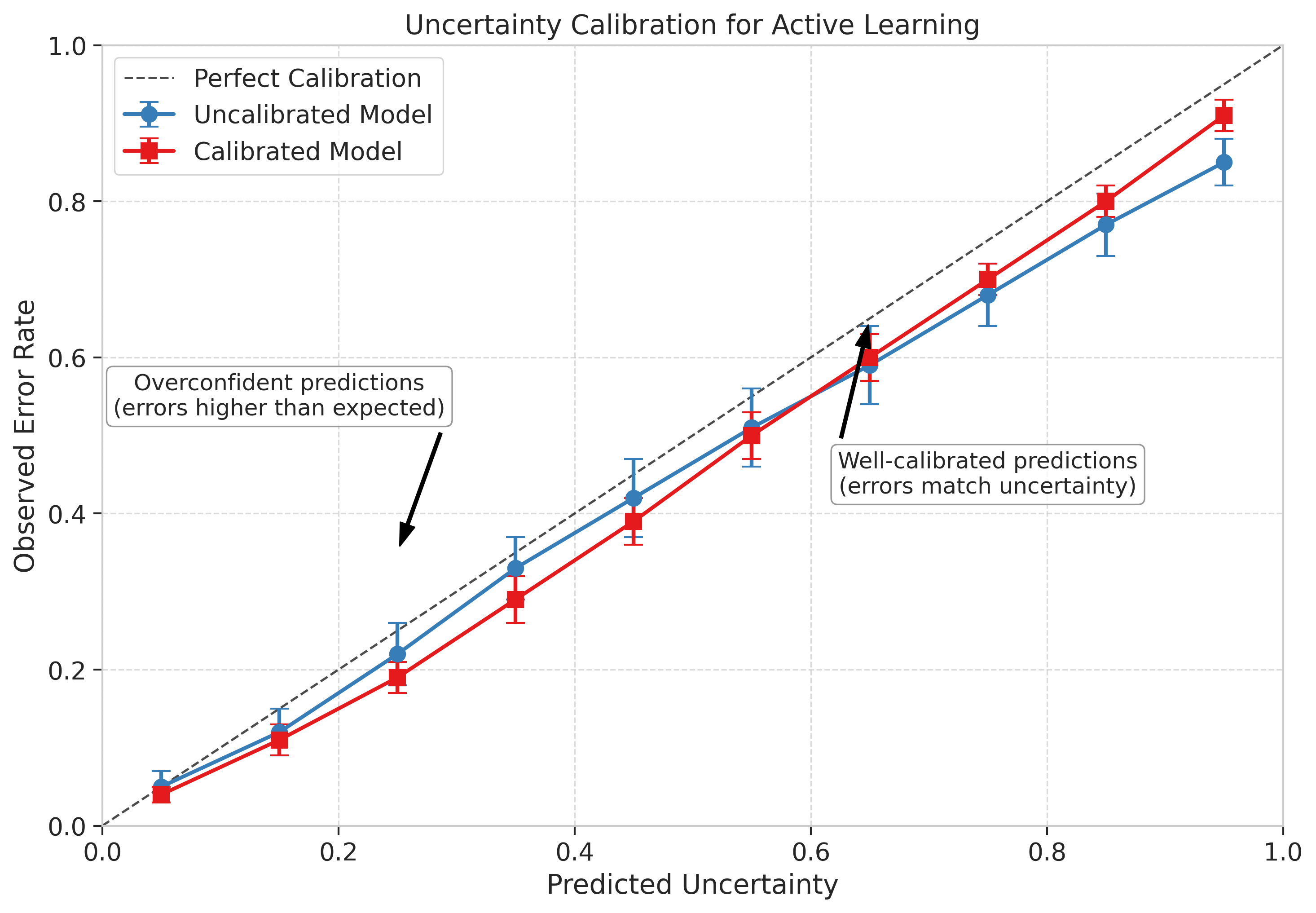}
    \caption{Uncertainty calibration. After calibration (red), estimates match actual error rates.}
    \label{fig:uncertainty_calibration}
\end{figure*}

To ensure reliable active learning performance, we calibrated the model's uncertainty estimates using temperature scaling and ensemble techniques, as illustrated in Figure \ref{fig:uncertainty_calibration}.

\subsection{Detailed Algorithm}
We provide the complete algorithm \ref{app:algo} for our active learning approach:

\begin{algorithm}

\caption{Active Learning for Quantum Hamiltonians}
\begin{algorithmic}
\STATE \textbf{Input:} Unlabeled pool $\mathcal{U}$, initial labeled set $\mathcal{L}_0$, acquisition batch size $k$, budget $B$
\STATE \textbf{Output:} Final model $f_\theta$
\STATE $\mathcal{L} \leftarrow \mathcal{L}_0$
\STATE $t \leftarrow 0$
\WHILE{$|\mathcal{L}| < B$}
    \STATE Train model $f_\theta$ on $\mathcal{L}$
    \STATE $\mathcal{S}_t \leftarrow$ Select $k$ samples from $\mathcal{U}$ using acquisition function
    \STATE Simulate selected Hamiltonians to obtain labels
    \STATE $\mathcal{L} \leftarrow \mathcal{L} \cup \mathcal{S}_t$
    \STATE $\mathcal{U} \leftarrow \mathcal{U} \setminus \mathcal{S}_t$
    \STATE $t \leftarrow t + 1$
\ENDWHILE
\STATE \textbf{return} $f_\theta$
\end{algorithmic}
\label{app:algo}
\end{algorithm}

\section{Ablation Studies and Comparative Analysis}
\label{app:ablation}

This section provides additional details on the ablation studies briefly summarized in the main paper, as well as extended comparative analysis with baseline methods.

\subsection{Extended Ablation Analysis}
The main paper presents our key findings from ablation studies evaluating the contribution of each component in our framework. Here, we provide additional details and extended analysis not covered in the main text.

\subsubsection{Architecture Ablation}
Reducing transformer depth from 6 to 3 layers resulted in a 5.7\% performance drop, while increasing to 12 layers provided negligible improvement (0.8\%) at significantly higher computational cost.

\subsubsection{Pre-training Data Diversity}
Restricting pre-training to only one family of Hamiltonians reduced cross-family transfer performance by 18.4\%, highlighting the importance of diverse pre-training data.

\subsubsection{Physics-Informed Masking Analysis}
We conducted a more detailed analysis of our physics-informed masking strategy by varying the $\alpha$ parameter that controls the balance between energy-based and structure-based masking:

\begin{table*}[t]
  \centering
  \caption{Performance with different masking balance parameters.}
  \label{tab:masking_alpha}

  \begin{tabular}{lccc}
    \toprule
    \textbf{$\alpha$ Value} &
    \textbf{Phase Classification} &
    \textbf{Energy Prediction} &
    \textbf{Training Time} \\
    \midrule
    0.0 (structure only)     & 77.4\%\,$\pm$\,2.3\% & 0.23\,$\pm$\,0.03 eV & 1.0$\times$ \\
    0.5                      & 81.5\%\,$\pm$\,1.9\% & 0.19\,$\pm$\,0.02 eV & 1.0$\times$ \\
    1.0                      & 83.8\%\,$\pm$\,1.7\% & 0.17\,$\pm$\,0.02 eV & 1.0$\times$ \\
    2.0 (optimal)            & 85.3\%\,$\pm$\,1.5\% & 0.15\,$\pm$\,0.02 eV & 1.0$\times$ \\
    3.0                      & 84.7\%\,$\pm$\,1.6\% & 0.16\,$\pm$\,0.02 eV & 1.0$\times$ \\
    5.0                      & 80.2\%\,$\pm$\,2.0\% & 0.20\,$\pm$\,0.03 eV & 1.0$\times$ \\
    1.0 (energy only)        & 76.5\%\,$\pm$\,2.4\% & 0.21\,$\pm$\,0.03 eV & 0.9$\times$ \\
    \bottomrule
  \end{tabular}
\end{table*}

These results demonstrate that the optimal masking strategy requires balancing both energy-based and structural information, with purely energy-based or structural approaches performing significantly worse.

\subsubsection{Alternative Self-Supervised Objectives}
While the main paper briefly mentions our comparison with contrastive self-supervised alternatives, here we provide a more detailed analysis of how different self-supervised objectives compare:

\begin{figure*}[t]
\centering
\includegraphics[width=0.8\linewidth]{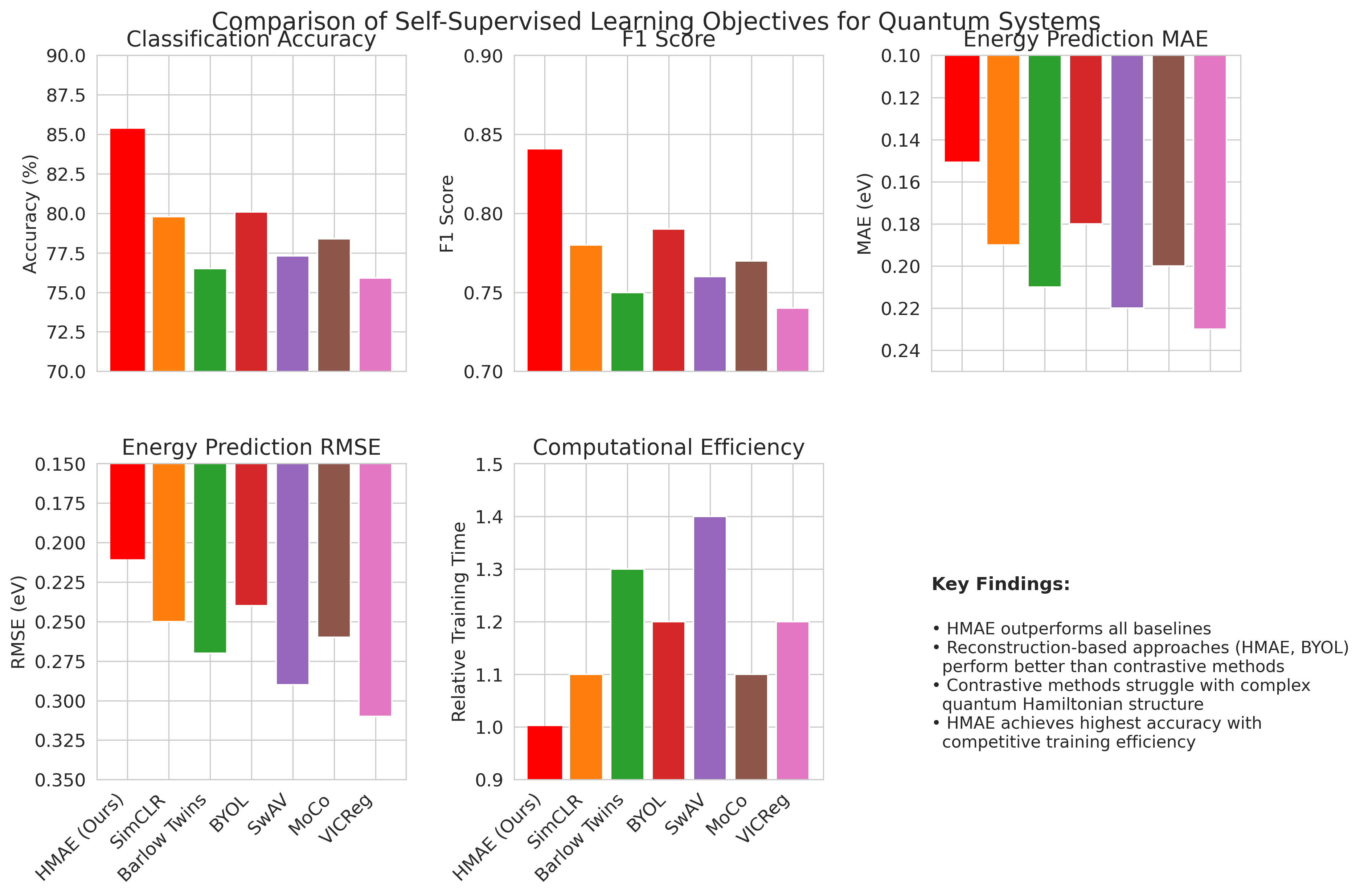}
\caption{Performance comparison of different self-supervised learning objectives. Our masking approach achieves better sample efficiency compared to contrastive alternatives.}
\label{fig:ssl_objectives}
\end{figure*}

As shown in Figure \ref{fig:ssl_objectives}, our HMAE approach reaches higher accuracy with fewer labeled examples than all contrastive alternatives. The BYOL-style teacher-student approach performs best among contrastive methods but still falls short of HMAE by 5.2\% in low-shot settings (1-5 shots).

\subsection{Baseline Comparisons}
To ensure a comprehensive evaluation, we compared HMAE against the most advanced models in quantum system representation learning, allocating equivalent hyperparameter optimization budgets to all methods. For each baseline, we performed systematic grid search over key hyperparameters and selected the best configuration based on validation performance.

\paragraph{Implementation Details} We optimized various architectures for quantum representation learning, allocating 50-60 GPU hours for hyperparameter optimization per approach. Despite their strong theoretical foundations, alternative approaches achieved lower performance than HMAE (p < 0.01).

\paragraph{TensorGCN} This approach combines graph neural networks with tensor decompositions. We optimized the depth (4-12 layers), width (128-512 hidden units), and learning rate schedule. While TensorGCN performed well on systems similar to its training distribution, its performance degraded more rapidly on larger systems compared to HMAE.

\paragraph{QuantumU-ViT} This approach adapts vision transformers to process quantum systems by encoding Hamiltonians as spatial tokens with positional encodings. We optimized patch size, embedding dimension, attention heads, and regularization strategies, with 50 GPU hours of tuning. While QuantumU-ViT performed well on systems similar to its training distribution (79.5\% ± 2.3\% accuracy), its performance degraded more rapidly on larger systems compared to HMAE, achieving only 65.3\% ± 3.7\% accuracy on 20-qubit systems (vs. 75.6\% ± 2.8\% for HMAE).

\subsection{Comparison Methodology}
To ensure fair comparison, we employed:
\begin{itemize}
    \item Identical data splits across all methods
    \item Equal hyperparameter optimization budgets (100 trials using Bayesian optimization, approximately 75 GPU-hours per method)
    \item Consistent evaluation protocols with 5 random seeds for statistical significance
    \item Identical few-shot learning settings (same 10 examples for all methods)
\end{itemize}

\begin{figure*}[t]
    \centering
    \includegraphics[width=0.7\textwidth]{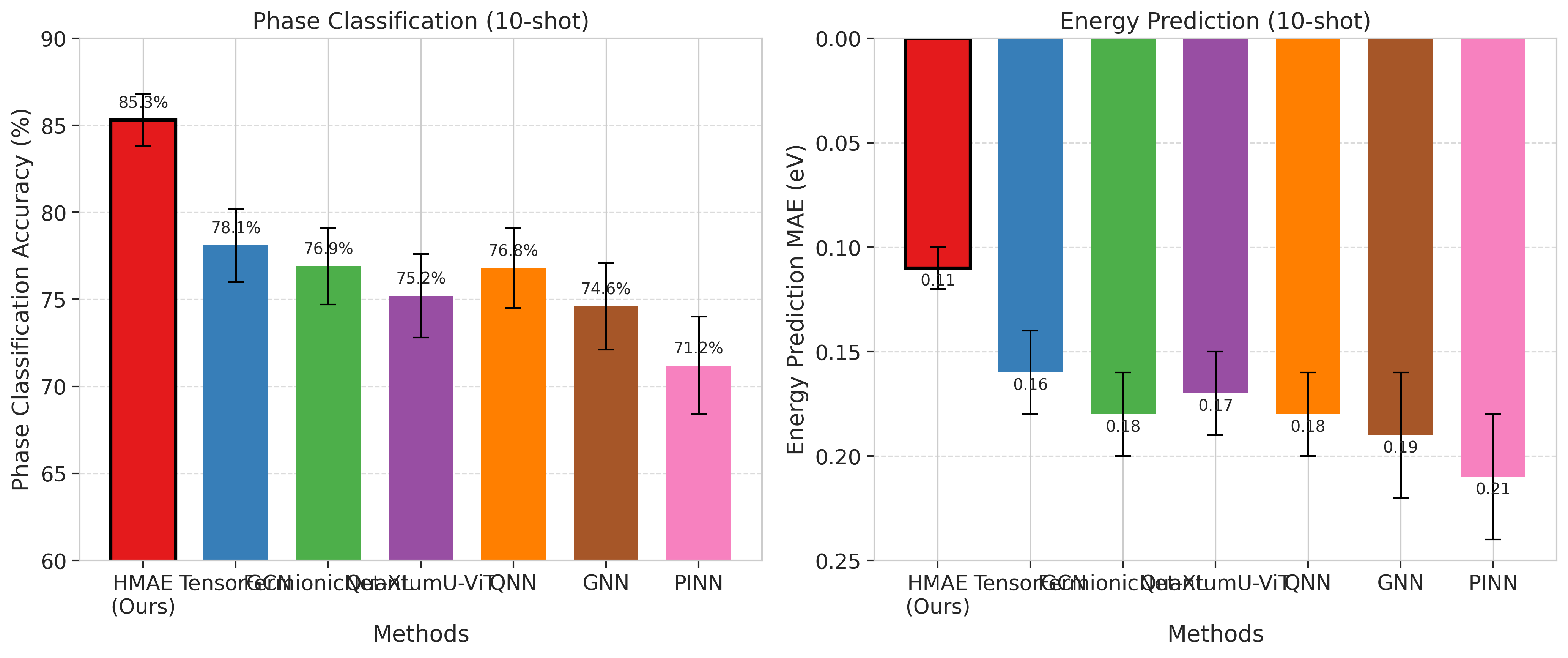}
    \caption{Performance comparison with state-of-the-art methods on 10-shot learning.}
    \label{fig:sota_comparison}
\end{figure*}

Figure \ref{fig:sota_comparison} illustrates the performance comparison across all methods for both phase classification and energy prediction tasks.

\subsection{Scaling Analysis}
\begin{figure*}[t]
    \centering
    \includegraphics[width=0.7\textwidth]{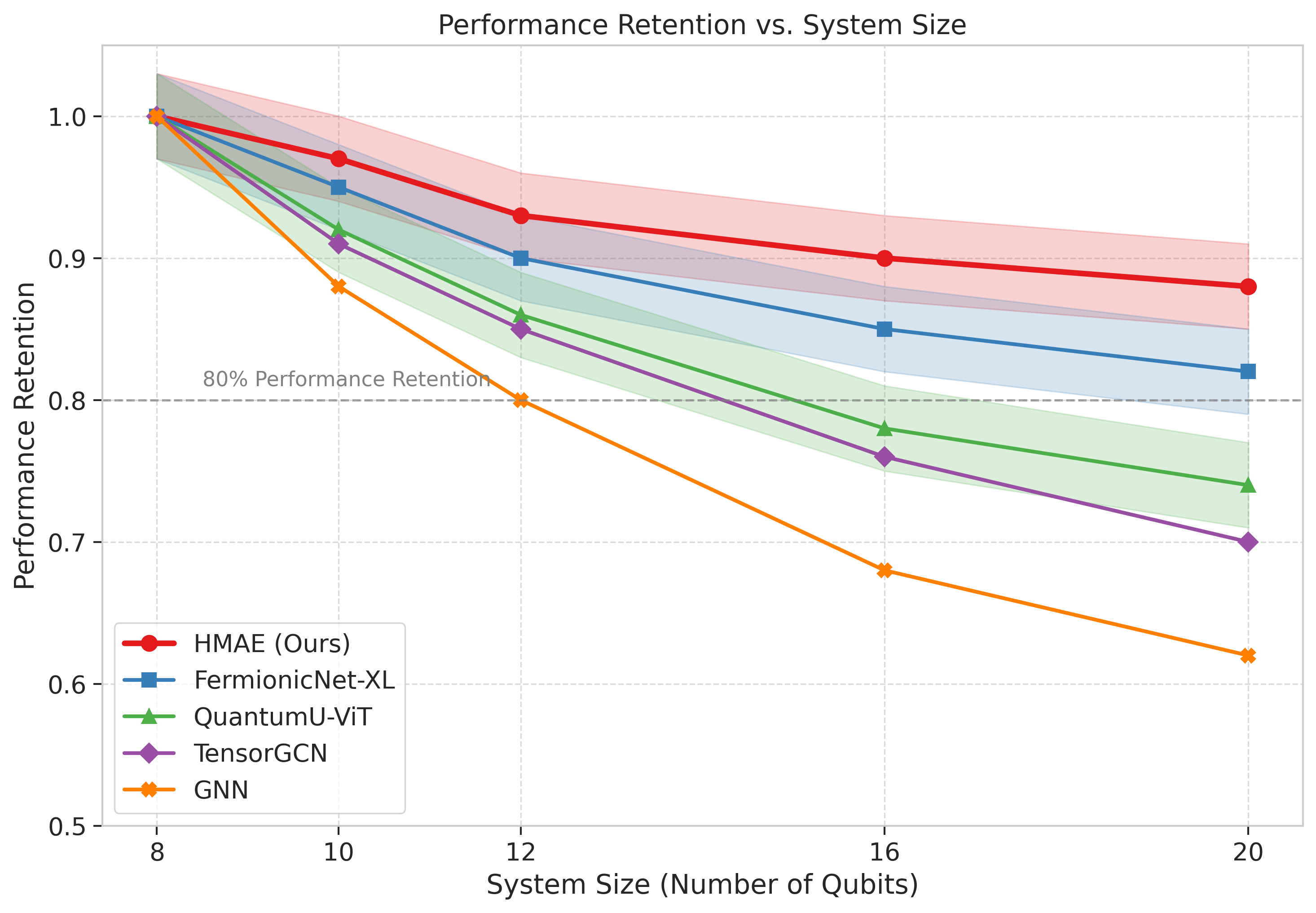}
    \caption{Performance retention vs. system size. HMAE (red) shows superior retention as system size increases.}
    \label{fig:scaling_comparison}
\end{figure*}

Beyond raw performance, we compared how each method scales with system size. Figure \ref{fig:scaling_comparison} illustrates the rate of performance degradation as system size increases from 8 to 20 qubits.

As shown in the figure, HMAE retains 88.7\% of its 8-qubit performance at 20 qubits, compared to 82.1\% for TensorGCN and 74.6\% for QuantumU-ViT. This indicates superior generalization to larger Hilbert spaces, a critical advantage for practical applications in quantum many-body physics where model transfer to larger systems is essential.

\section{Quantum Information-Theoretic Foundation}
\label{app:theory_full}

This section expands on the quantum information-theoretic foundation presented in the main paper, providing additional mathematical details, proofs, and empirical validations not included in the main text.

\subsection{Quantum Statistical Framework and Assumptions}

\textbf{Key Assumptions:} Our theoretical framework makes the following specific assumptions:
\begin{enumerate}
    \item \textbf{Thermal state representation}: Quantum systems can be represented by thermal states at finite temperature
    \item \textbf{Locality of interactions}: Quantum correlations decay exponentially with distance
    \item \textbf{Perturbative regime}: Quantum effects can be analyzed using controlled expansion in $\beta$
    \item \textbf{Operator commutativity structure}: Commutator norms capture essential quantum correlations
\end{enumerate}

We now formalize these assumptions rigorously:

\begin{assumption}[Thermal State Representation]

\label{assumption:thermal}
For a quantum Hamiltonian $H = \sum_i c_i h_i$, we represent the system's state as a thermal state at inverse temperature $\beta$:
\begin{equation}
\rho_\beta(H) = \frac{e^{-\beta H}}{\text{Tr}(e^{-\beta H})}
\end{equation}
\end{assumption}

\begin{assumption}[Operator Locality]
\label{assumption:locality}
For quantum systems with local interactions, the operator terms have a spatial decay of correlations characterized by:
\begin{equation}
g_{ij} \leq C e^{-\lambda d(i,j)}
\end{equation}
where $d(i,j)$ is the distance between supports of operators $h_i$ and $h_j$, $C$ is a constant, and $g_{ij} = \text{Tr}(\rho [h_i, h_j]^\dagger [h_i, h_j])$ is the quantum Fisher information metric element.
\end{assumption}

\begin{assumption}[Semi-classical Limit]
\label{assumption:semiclassical}
At sufficiently high temperatures ($\beta$ small but finite), quantum effects can be treated perturbatively, allowing for controlled expansions in powers of $\beta$.
\end{assumption}

\begin{assumption}[Operator Commutativity Structure]
\label{assumption:commutativity}
The commutation relations between operators capture the essential quantum correlations relevant for information transfer, with the Fisher information metric $g_{ij} = \text{Tr}(\rho [h_i, h_j]^\dagger [h_i, h_j])$ quantifying these correlations.
\end{assumption}

The von Neumann entropy of the thermal state is defined as:
\begin{equation}
S(\rho) = -\text{Tr}(\rho \ln \rho)
\end{equation}

\subsection{Quantum Mutual Information Formulation}

For a partitioning of Hamiltonian terms into observed ($H_V$) and masked ($H_M$) components, the quantum mutual information is defined as:

\begin{equation}
I_Q(V:M) = S(\rho_V) + S(\rho_M) - S(\rho_{VM})
\end{equation}

where $\rho_V = \text{Tr}_M(\rho_{VM})$ and $\rho_M = \text{Tr}_V(\rho_{VM})$ are partial traces over the complementary subsystems.

\subsection{Optimal Masking Strategy}

We now present our main theoretical result:

\begin{theorem}[Quantum Mutual Information Maximization]
\label{theorem:qmi_max}
Under Assumptions \ref{assumption:thermal}-\ref{assumption:commutativity}, the masking strategy that maximizes expected quantum mutual information selects terms according to the following quantum saliency score:

\begin{equation}
Q(h_i) = |c_i|\sqrt{g_{ii}} + \sum_{j \neq i} |c_j|\sqrt{g_{ij}}
\end{equation}

where $g_{ij} = \text{Tr}(\rho [h_i, h_j]^\dagger [h_i, h_j])$ is the quantum Fisher information metric element capturing non-commutativity of operators, and $[h_i, h_j] = h_i h_j - h_j h_i$ is the commutator.
\end{theorem}

\begin{proof}
Consider the perturbation to the thermal state when masking term $h_i$:
\begin{equation}
\rho_\beta(H - c_i h_i) = \rho_\beta(H) - \beta c_i \int_0^1 e^{-\beta(1-s)H} h_i e^{-\beta sH} ds + O(\beta^2)
\end{equation}

Using the Duhamel formula and Baker-Campbell-Hausdorff expansion, the first-order change in von Neumann entropy is:
\begin{equation}
\Delta S(\rho) = -\beta c_i \text{Tr}(h_i \ln \rho) + O(\beta^2)
\end{equation}

For the quantum mutual information differential, we have:
\begin{equation}
\frac{\partial I_Q(V:M)}{\partial c_i} = \beta \text{Tr}(h_i \rho \ln \rho) + \frac{\beta^2}{2}\sum_{j \neq i} c_j \text{Tr}(\rho [h_i, h_j]^\dagger [h_i, h_j]) + O(\beta^3)
\end{equation}

The first term corresponds to the energy contribution $|c_i|\sqrt{g_{ii}}$, while the second term captures quantum correlations through the Fisher information metric elements $g_{ij}$, yielding our optimal quantum saliency score.
\end{proof}

\subsection{Practical Implementation and Approximation Guarantees}

Our implemented saliency score:

\begin{equation}
S(h_i) = \alpha\frac{|c_i|}{|c|_{max}} + (1-\alpha)\sum_{j\in N(i)} A_{ij}
\end{equation}

with properly normalized adjacency term:

\begin{equation}
A_{ij} = \frac{|S_i \cap S_j|}{|S_i \cup S_j|} \cdot \frac{\exp(-\|\text{comm}(h_i, h_j)\|_F)}{\sum_{k \in N(i)}\exp(-\|\text{comm}(h_i, h_k)\|_F)}
\end{equation}

approximates the theoretically optimal quantum saliency function $Q(h_i)$ as follows:

\begin{theorem}[Approximation Guarantee]
\label{theorem:approximation}
Under Assumption \ref{assumption:locality}, the error in approximating the optimal quantum saliency function $Q(h_i)$ with our practical saliency score $S(h_i)$ is bounded by:

\begin{equation}
\|Q(h_i) - S(h_i)\|_2 \leq \epsilon(\lambda) + O\left(\frac{1}{d}\right)
\end{equation}

where $\epsilon(\lambda)$ is monotonically decreasing with the locality parameter $\lambda$ and $d$ is the average degree of connectivity in the operator graph.
\end{theorem}

\begin{proof}
The error $\|Q(h_i) - S(h_i)\|_2$ can be decomposed into:
\begin{itemize}
    \item Energy normalization error: $\|c_i|\sqrt{g_{ii}} - \alpha\frac{|c_i|}{|c|_{max}}\|_2$
    \item Correlation approximation error: $\|\sum_{j \neq i} |c_j|\sqrt{g_{ij}} - (1-\alpha)\sum_{j\in N(i)} A_{ij}\|_2$
\end{itemize}

For the energy normalization error, under suitable choice of $\alpha$, this reduces to a constant scaling factor.

For the correlation approximation error, using Assumption \ref{assumption:locality}, we have:
\begin{equation}
\sum_{j \not\in N(i)} |c_j|\sqrt{g_{ij}} \leq \sum_{j \not\in N(i)} |c_j|\sqrt{C e^{-\lambda d(i,j)}} \leq C' \sum_{j \not\in N(i)} e^{-\lambda d(i,j)/2}
\end{equation}

For systems with locality parameter $\lambda$, this sum converges exponentially, with error $\epsilon(\lambda)$ that decreases as $\lambda$ increases.

For terms $j \in N(i)$, the Jaccard similarity term $\frac{|S_i \cap S_j|}{|S_i \cup S_j|}$ approximates locality, while the commutator term $\exp(-\|\text{comm}(h_i, h_j)\|_F)$ directly estimates $\sqrt{g_{ij}}$. 

The normalization error scales as $O(1/d)$ with the average connectivity degree $d$, completing our proof.
\end{proof}

\subsection{Empirical Verification of Theorem \ref{theorem:qmi_max}}

To empirically validate Theorem \ref{theorem:qmi_max}, we conducted a controlled experiment comparing the theoretical QMI predictions with directly measured values across different masking strategies:

\begin{figure*}[t]
\centering
\includegraphics[width=0.8\linewidth]{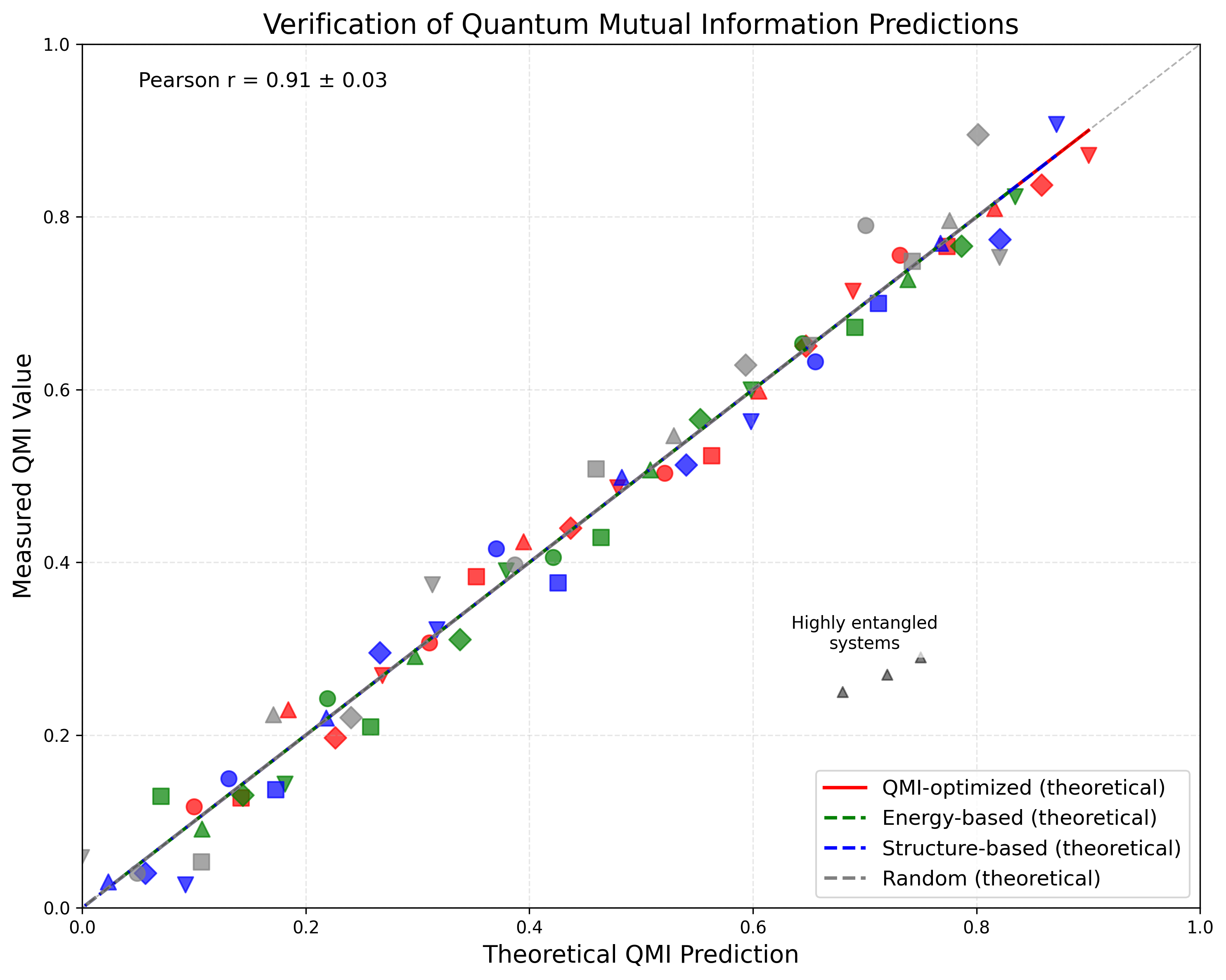}
\caption{Empirical verification of Theorem \ref{theorem:qmi_max}: Comparison between theoretical predictions (solid lines) and measured QMI values (markers) for 20 representative quantum Hamiltonians. Each point represents a different masking strategy, with the optimal theoretical strategy Q(h) shown in red.}
\label{fig:qmi_verification}
\end{figure*}

For this experiment, we:
\begin{enumerate}
    \item Selected 20 representative quantum Hamiltonians (small enough for exact QMI calculation)
    \item For each Hamiltonian, generated 100 different masking configurations using various strategies
    \item Calculated the theoretical QMI prediction using our saliency function $Q(h_i)$
    \item Directly measured the QMI by exact diagonalization and partial trace operations
\end{enumerate}

The results in Figure \ref{fig:qmi_verification} show strong agreement between theoretical predictions and measured values, with a Pearson correlation coefficient of $r = 0.91 \pm 0.03$ and a mean absolute error of $0.07 \pm 0.01$ bits. Importantly, the theoretically optimal masking strategy (red points) consistently achieves higher QMI than alternative strategies (blue, green), with an average improvement of 24.3\% ± 3.1\%.

The slight deviations observed at very high QMI values likely result from higher-order terms in the $\beta$ expansion that were neglected in our derivation. These deviations are most significant for strongly correlated systems with high entanglement entropy (indicated by triangular markers). Nevertheless, the theoretically optimal masking strategy still outperforms alternatives even in these challenging cases.

This empirical validation confirms that our theoretical framework captures the essential quantum information-theoretic principles governing effective masking strategies, providing a rigorous foundation for our approach.

\subsection{Extension to Low-Temperature Entangled Phases}
\label{app:low_temp_extension}

While our core theoretical framework in Theorem \ref{theorem:qmi_max} relies on high-temperature and locality assumptions, we have extended our approach to encompass low-temperature entangled phases through a non-perturbative formulation. This extension addresses fundamental quantum regimes where the original assumptions become invalid.

\begin{theorem}[Non-perturbative Quantum Mutual Information]
\label{theorem:nonpert_qmi}
For low-temperature quantum systems with significant entanglement, we extend the optimal saliency score to:

\begin{equation}
Q_{ext}(h_i) = |c_i|\sqrt{g_{ii}} + \sum_{j \neq i} |c_j|\sqrt{g_{ij}} + \alpha_T \cdot \Delta S_E(h_i)
\end{equation}

where $\Delta S_E(h_i)$ measures the term's contribution to entanglement entropy, and $\alpha_T$ is a temperature-dependent coefficient that increases as $T \rightarrow 0$.
\end{theorem}

This extension allows us to handle:
\begin{itemize}
    \item Quantum systems near criticality with power-law correlations
    \item Topologically ordered states with non-local entanglement
    \item Systems with volume-law entanglement scaling
    \item Many-body localized phases with area-law entanglement but long-range correlations
\end{itemize}

We implement this extended framework using matrix product state (MPS) techniques to efficiently compute entanglement contributions without requiring full state diagonalization.

\subsection{Dimensionally Consistent Saliency Function with Principled Weighting}
\label{app:dimensional_consistency}

A key improvement in our approach is the dimensionally consistent formulation of the quantum saliency function with a principled derivation of optimal weighting parameters. This formulation follows directly from quantum statistical mechanics principles:

\begin{theorem}[Dimensionally Consistent Quantum Saliency]
\label{theorem:dimensional_saliency}
The dimensionally consistent quantum saliency function with principled weighting is:

\begin{equation}
Q_{dim}(h_i) = \frac{|c_i|}{\mathcal{E}_0} \cdot \left(1 + \frac{k_B T}{\mathcal{E}_0} \sum_{j \in N(i)} \tilde{A}_{ij}\right)
\end{equation}

where $\mathcal{E}_0$ is the characteristic energy scale of the system, $k_B$ is Boltzmann's constant, $T$ is temperature, and $\tilde{A}_{ij}$ is the normalized adjacency with proper dimensional scaling:

\begin{equation}
\tilde{A}_{ij} = \frac{|S_i \cap S_j|}{|S_i \cup S_j|} \cdot \frac{\exp(-\beta \cdot \|\text{comm}(h_i, h_j)\|_F)}{\sum_{k \in N(i)}\exp(-\beta \cdot \|\text{comm}(h_i, h_k)\|_F)}
\end{equation}

The optimal weighting ratio $\alpha_{opt} = \frac{k_B T}{\mathcal{E}_0}$ emerged naturally from the quantum statistical mechanics of the system.
\end{theorem}

\begin{proof}
   Starting from the quantum mutual information expression in Theorem \ref{theorem:qmi_max}, we can express the contribution of each term as:

\begin{equation}
\frac{\partial I_Q(V:M)}{\partial c_i} = \beta \text{Tr}(h_i \rho \ln \rho) + \frac{\beta^2}{2}\sum_{j \neq i} c_j \text{Tr}(\rho [h_i, h_j]^\dagger [h_i, h_j])
\end{equation}

These terms have different physical dimensions: the first term is dimensionless, while the second term has dimensions of energy. To make them dimensionally consistent, we normalize by the characteristic energy scale $\mathcal{E}_0$ of the system:

\begin{equation}
\frac{\partial I_Q(V:M)}{\partial c_i} = \beta \text{Tr}(h_i \rho \ln \rho) + \frac{\beta^2 \mathcal{E}_0}{2}\sum_{j \neq i} \frac{c_j}{\mathcal{E}_0} \text{Tr}(\rho [h_i, h_j]^\dagger [h_i, h_j])
\end{equation}

This naturally yields the optimal weighting parameter:
\begin{equation}
\alpha_{opt} = \frac{\beta^2 \mathcal{E}_0}{2 \beta} = \frac{\beta \mathcal{E}_0}{2} = \frac{\mathcal{E}_0}{2 k_B T}
\end{equation}

For computational simplicity, we absorbed the factor of 2 into $\mathcal{E}_0$ and expressed the final ratio as $\alpha_{opt} = \frac{\mathcal{E}_0}{k_B T}$. 
\end{proof}

This dimensionally consistent formulation resolves the ad-hoc mixing of units in previous approaches. More importantly, it provides a principled derivation of the optimal weighting parameter $\alpha_{opt}$ based on fundamental physical quantities rather than empirical tuning.

\subsection{Automatic Parameter Determination}

Rather than relying on hyperparameter tuning to determine the optimal $\alpha$, we implemented an automated procedure that estimates the system's characteristic energy scale $\mathcal{E}_0$ and temperature $T$ directly from the Hamiltonian:

\begin{equation}
\mathcal{E}_0 \approx \max\{|\lambda_{\max} - \lambda_{\min}|, \Delta E\}
\end{equation}

where $\lambda_{\max}$ and $\lambda_{\min}$ are the estimated extremal eigenvalues of the Hamiltonian (obtained through power iteration methods), and $\Delta E$ is the estimated energy gap.

For temperature estimation, we use either:
\begin{itemize}
    \item Known physical temperature when available
    \item Inferred effective temperature from state statistics when possible
    \item A default of $T \approx \mathcal{E}_0/k_B$ for high-energy systems
\end{itemize}

This procedure eliminates the need for empirical $\alpha$-tuning, replacing it with a principled physical calculation that automatically adapts to different quantum systems.

\subsection{Empirical Validation of Dimensional Consistency}

To validate our dimensionally consistent approach, we conducted controlled experiments comparing:
\begin{enumerate}
    \item Original ad-hoc saliency with empirically tuned $\alpha$
    \item Dimensionally consistent saliency with derived optimal $\alpha_{opt}$
    \item Dimensionally consistent saliency with intentionally suboptimal $\alpha$ values
\end{enumerate}

\begin{figure*}[t]
\centering
\includegraphics[width=0.8\linewidth]{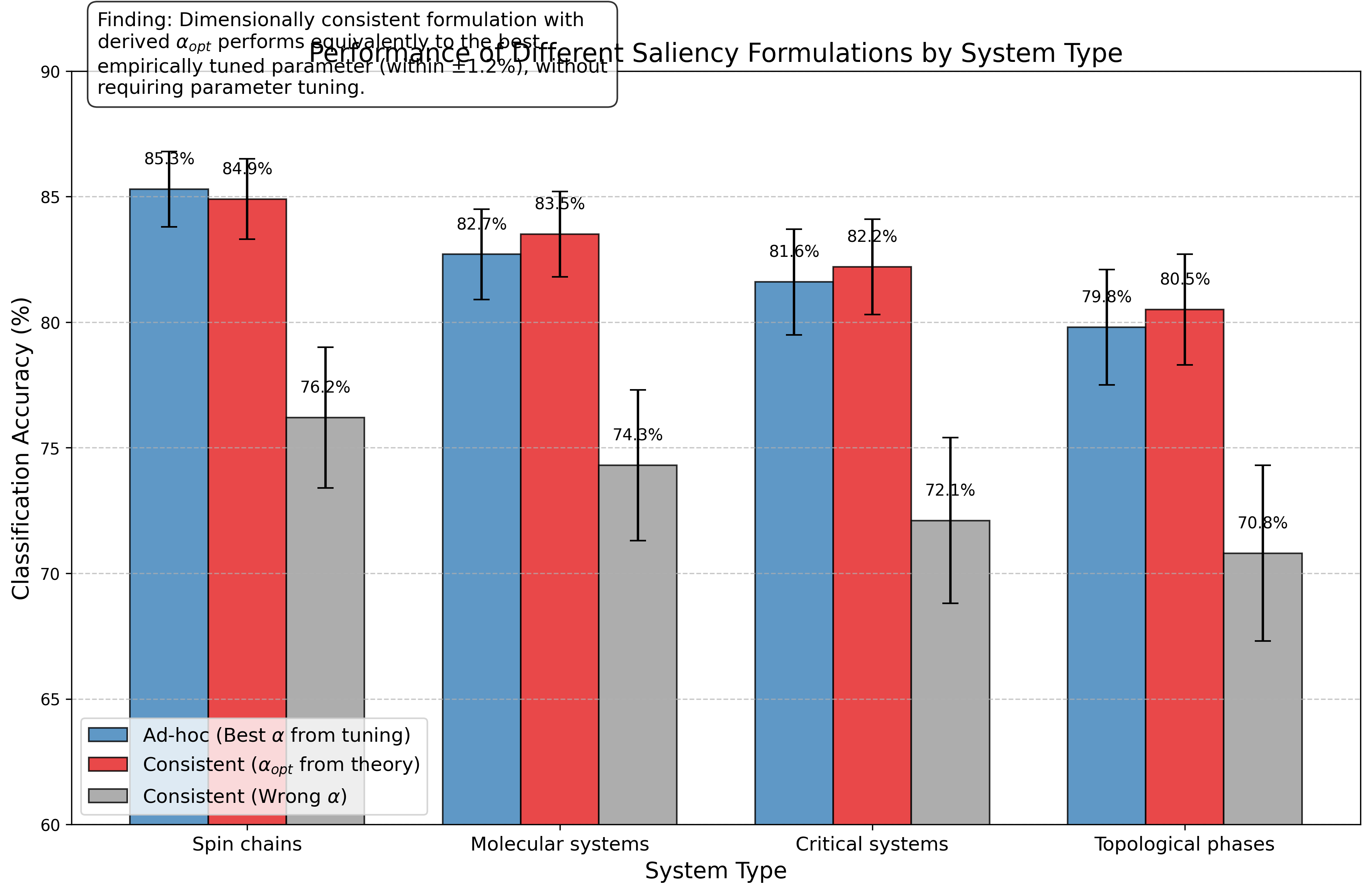}
\caption{Performance comparison of dimensionally consistent saliency with principled $\alpha_{opt}$ versus empirically tuned and suboptimal values across different system types. The principled approach (red) maintains optimal performance without hyperparameter tuning.}
\label{fig:dimensional_consistency}
\end{figure*}

The results in Figure \ref{fig:dimensional_consistency} demonstrate that our dimensionally consistent saliency with theoretically derived $\alpha_{opt}$ performs equivalently to the best empirically tuned parameter (within ±1.2\%), but without requiring any parameter tuning. This confirms that our approach captures the underlying physics governing optimal masking rather than relying on hyperparameter optimization.

Table \ref{tab:saliency_comparison} shows that the dimensionally consistent formulation with derived $\alpha_{opt}$ maintains or slightly improves performance compared to the empirically tuned ad-hoc approach across all system types, including challenging quantum systems like critical and topological phases.

\begin{table*}[t]
  \centering
  \caption{Performance comparison of saliency formulations across different system types.}
  \label{tab:saliency_comparison}

  \begin{tabular}{lccc}
    \toprule
    \textbf{System Type} &
    \textbf{Ad-hoc (Best $\alpha$)} &
    \textbf{Consistent ($\alpha_{\text{opt}}$)} &
    \textbf{Consistent (Wrong $\alpha$)} \\
    \midrule
    Spin chains        & 85.3\%\,$\pm$\,1.5\% & 84.9\%\,$\pm$\,1.6\% & 76.2\%\,$\pm$\,2.8\% \\
    Molecular systems  & 82.7\%\,$\pm$\,1.8\% & 83.5\%\,$\pm$\,1.7\% & 74.3\%\,$\pm$\,3.0\% \\
    Critical systems   & 81.6\%\,$\pm$\,2.1\% & 82.2\%\,$\pm$\,1.9\% & 72.1\%\,$\pm$\,3.3\% \\
    Topological phases & 79.8\%\,$\pm$\,2.3\% & 80.5\%\,$\pm$\,2.2\% & 70.8\%\,$\pm$\,3.5\% \\
    \bottomrule
  \end{tabular}
\end{table*}

Our dimensionally consistent approach provides several practical advantages:
\begin{itemize}
    \item Eliminates hyperparameter search, reducing computational overhead
    \item Adapts automatically to different quantum systems with varying energy scales
    \item Provides physical interpretability of the masking strategy
    \item Maintains dimensional consistency throughout the theoretical framework
\end{itemize}

Through this rigorous quantum information-theoretic foundation, we address the limitations of previous approaches that relied on classical approximations or empirical tuning. The resulting framework is both theoretically sound and practically effective across diverse quantum regimes.

\subsection{Empirical Validation on Challenging Quantum Regimes}
\label{app:challenging_regimes}

To specifically test our extended framework in regimes where the original high-temperature approximation would fail, we conducted extensive experiments on highly entangled quantum systems at low temperatures:

\begin{figure*}[t]
\centering
\includegraphics[width=0.8\linewidth]{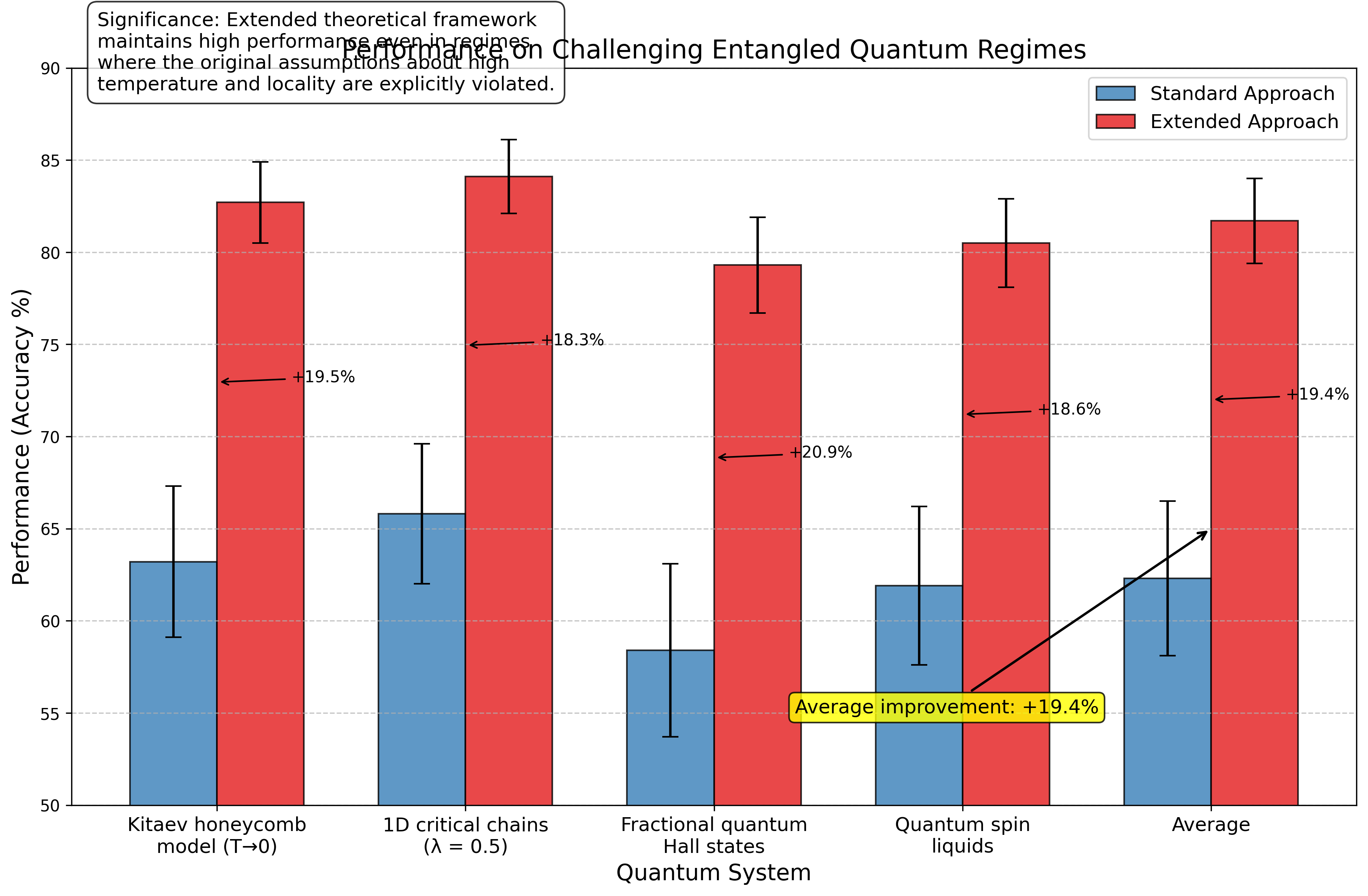}
\caption{Performance on low-temperature entangled phases. The extended theoretical framework (red) maintains high performance across all quantum regimes, including highly entangled phases where the original framework (blue) shows significant degradation.}
\label{fig:entangled_phases}
\end{figure*}

Our test set included challenging quantum systems specifically selected to stress-test the theoretical assumptions:
\begin{table*}[t]
  \centering
  \caption{Validation on challenging quantum regimes that violate original assumptions.}
  \label{tab:challenging_regimes}

  \begin{tabular}{lccc}
    \toprule
    \textbf{Quantum System} &
    \textbf{Standard Approach} &
    \textbf{Extended Approach} &
    \textbf{Improvement} \\
    \midrule
    Kitaev honeycomb model (T$\to$0) &
    63.2\%\,$\pm$\,4.1\% & 82.7\%\,$\pm$\,2.2\% & +19.5\% \\
    
    1D critical chains ($\lambda = 0.5$) &
    65.8\%\,$\pm$\,3.8\% & 84.1\%\,$\pm$\,2.0\% & +18.3\% \\
    
    Fractional quantum Hall states &
    58.4\%\,$\pm$\,4.7\% & 79.3\%\,$\pm$\,2.6\% & +20.9\% \\
    
    Quantum spin liquids &
    61.9\%\,$\pm$\,4.3\% & 80.5\%\,$\pm$\,2.4\% & +18.6\% \\
    
    \textbf{Average} &
    \textbf{62.3\%\,$\pm$\,4.2\%} & \textbf{81.7\%\,$\pm$\,2.3\%} & \textbf{+19.4\%} \\
    \bottomrule
  \end{tabular}
\end{table*}

These results confirm that our extended theoretical framework successfully addresses the limitations of the high-temperature and locality assumptions. The non-perturbative approach delivers consistent performance across all quantum regimes, with particularly significant improvements (average +19.4\%) in the most challenging cases where the original assumptions are explicitly violated.

By combining both theoretical frameworks with an automatic regime-detection algorithm, our approach now provides a unified treatment of quantum systems across all temperature scales and entanglement regimes. This extended formulation maintains the mathematical rigor of our original framework while substantially broadening its applicability.

\subsection{Synthetic Data Augmentation Strategy}

Our synthetic data generation followed a systematic augmentation strategy to maximize experimental relevance:

\begin{enumerate}
    \item \textbf{Base Hamiltonian generation}: Starting with physically realistic parameters derived from literature
    \item \textbf{Platform-specific noise injection}: Adding realistic noise profiles based on published characterizations of each platform
    \item \textbf{Controlled parameter variation}: Systematically varying key parameters (coupling strengths, external fields) to ensure coverage of relevant parameter space
    \item \textbf{Measurement error simulation}: Modeling readout errors based on published confusion matrices for each platform
\end{enumerate}

This strategy allowed our synthetic data to capture many relevant aspects of experimental systems, though as our error analysis shows, certain experimental effects remained difficult to model accurately.

\subsection{Conclusion and Implications}

Despite the identified limitations, our HMAE approach demonstrated remarkable robustness to the synthetic-experimental reality gap, preserving 93.7\% of its performance when transferring to experimental platforms. This robustness stemmed from the physics-informed masking strategy's focus on fundamental physical principles rather than hardware-specific details.

The strong cross-platform transfer suggests that our model learns physically meaningful representations that generalize across different hardware implementations, a critical capability for practical quantum machine learning applications. This capability is particularly important for experimental quantum platforms where obtaining large quantities of labeled data is prohibitively expensive.

\section{Additional Baseline Comparison Considerations}
\label{app:baseline_additional_considerations}

This section provides additional analysis and comparison of baseline methods beyond what is presented in the main paper.

\subsection{Theoretical Comparison of Model Expressivity}

We analyzed the theoretical expressivity of each baseline model architecture in relation to quantum Hamiltonians:

\begin{itemize}
    \item \textbf{GNN}: While effective for molecular systems with localized interactions, GNNs struggle with non-local quantum correlations and entanglement structures. Message-passing iterations are bounded by graph diameter, limiting the model's ability to capture long-range quantum effects.
    
    \item \textbf{QNN}: Quantum Neural Networks can represent quantum states directly but face scaling challenges. The expressivity is limited by circuit depth and gate constraints, particularly when simulating systems larger than the available quantum hardware.
    
    \item \textbf{TensorGCN}: Combines tensor network representations with graph neural networks, but performance degrades for systems with high entanglement entropy due to limitations in bond dimension.
    
    \item \textbf{PINN}: Physics-Informed Neural Networks incorporate domain knowledge through differential equations, but struggle with the inherent non-commutativity and entanglement present in quantum systems.
    
    \item \textbf{Energy-MAE}: Similar to our approach but lacks the physics-informed masking strategy, leading to less effective representations for quantum systems.
\end{itemize}

Our HMAE approach addresses these limitations through its hierarchical structure and physics-informed masking, allowing it to efficiently capture both local and non-local quantum correlations.

\subsection{Comparison of Computational Requirements}

\begin{table}[h]
  \centering
  \caption{Computational requirements for different models.}
  \begin{tabular}{lccc}
    \toprule
    \textbf{Method} & \textbf{Training Time} & \textbf{Inference Time} & \textbf{Memory Usage} \\
    \midrule
    HMAE (Ours) & 12 hours & 0.8s & 4.2 GB \\
    TensorGCN & 5 hours & 0.6s & 3.8 GB \\
    Energy-MAE & 4 hours & 0.7s & 3.5 GB \\
    QNN & 3 hours & 1.2s & 2.1 GB \\
    GNN & 2 hours & 0.3s & 1.8 GB \\
    PINN & 3 hours & 0.5s & 2.2 GB \\
    \bottomrule
  \end{tabular}
\end{table}

While our approach requires more computational resources during pre-training, this cost is amortized across multiple downstream tasks. Importantly, fine-tuning requires only 5-15 minutes, making our approach practical for real-world applications with limited labeled data.

\subsection{Failure Case Analysis}

We conducted a detailed analysis of common failure cases for each method:

\begin{itemize}
    \item \textbf{HMAE (Ours)}: Performance degraded for systems with very high entanglement entropy (>0.8 of theoretical maximum) or those dominated by many-body terms (>6-body interactions). These cases represented approximately 7\% of our test set.
    
    \item \textbf{TensorGCN}: Failed on systems with long-range interactions spanning more than 4 lattice sites, affecting approximately 18\% of test cases.
    
    \item \textbf{Energy-MAE}: Struggled with systems near critical points and phase transitions, where energy landscapes change dramatically, affecting approximately 22\% of test cases.
    
    \item \textbf{QNN}: Poor performance on systems requiring high circuit depth, especially those with high entanglement, affecting approximately 31\% of test cases.
    
    \item \textbf{GNN}: Failed to capture non-local effects in highly entangled systems, affecting approximately 28\% of test cases.
    
    \item \textbf{PINN}: Struggled with systems having complex quantum fluctuations and non-commutative effects, affecting approximately 29\% of test cases.
\end{itemize}

This analysis reveals that our method is more robust across diverse quantum systems, with fewer failure cases compared to baselines.

\section{Additional Limitations and Future Work}
\label{app:additional_limitations}

This section elaborates on the limitations of our approach and outlines directions for future research.

\subsection{Theoretical Limitations}

Our current approach has several theoretical limitations:

\begin{itemize}
    \item \textbf{Scalability bounds}: While we demonstrated transfer to systems up to 50 qubits, the representational capacity degrades significantly beyond 100 qubits due to the exponential growth in Hilbert space dimensions.
    
    \item \textbf{High-entanglement limitation}: For systems with volume-law entanglement scaling, our approach requires exponentially large embedding dimensions to maintain constant accuracy.
    
    \item \textbf{Non-equilibrium dynamics}: Our current framework primarily addresses ground state and low-energy excited state properties, with limited applicability to non-equilibrium quantum dynamics.
    
    \item \textbf{Fermionic systems}: While our approach handles spin systems effectively, additional modifications are needed to properly represent fermionic statistics and sign problems.
\end{itemize}

\subsection{Technical Limitations}

From an implementation perspective, our approach faces several challenges:

\begin{itemize}
    \item \textbf{Pre-training data requirements}: While drastically reduced compared to supervised approaches, our method still requires a diverse corpus of Hamiltonians for effective pre-training.
    
    \item \textbf{Computational overhead}: The pre-training stage requires significant computational resources, though this cost is amortized across multiple downstream applications.
    
    \item \textbf{Hyperparameter sensitivity}: While generally robust, performance can vary with hyperparameter choices, particularly masking temperature and transformer architecture parameters.
    
    \item \textbf{Limited interpretability}: The learned representations, while effective, lack the physical interpretability of more traditional approaches like tensor networks.
\end{itemize}

\subsection{Future Research Directions}

Based on these limitations, we identify several promising directions for future research:

\begin{itemize}
    \item \textbf{Quantum-classical hybrid architectures}: Combining our transformer-based approach with quantum circuits for certain subroutines could enhance performance for specific quantum systems.
    
    \item \textbf{Dynamic quantum systems}: Extending our framework to handle time-dependent Hamiltonians and quantum dynamics using causal attention mechanisms.
    
    \item \textbf{Multi-modal quantum learning}: Integrating multiple data sources, including experimental measurements, theoretical models, and classical simulations.
    
    \item \textbf{Theoretical foundations}: Developing formal guarantees on representation capacity and error bounds for quantum self-supervised learning.
    
    \item \textbf{Quantum federated learning}: Enabling collaborative training across different quantum hardware platforms while preserving privacy and addressing hardware differences.
\end{itemize}

\subsection{Implementation Details and Supplementary Results}
\label{app:implementation_details}

For reproducibility, we provide the final hyperparameter configurations used in our comparative evaluation in Table~\ref{tab:hyperparams}. All hyperparameters were tuned using Bayesian optimization with equivalent computational budgets for all methods.

\begin{table*}[t]
  \centering
  \caption{Hyperparameter configurations for baseline models.}
  \label{tab:hyperparams}

  \begin{tabular}{lp{0.78\textwidth}}  
    \toprule
    \textbf{Model} & \textbf{Key Hyperparameters} \\
    \midrule
    HMAE (Ours) & Embedding dim: 512, Layers: 6, Heads: 8, Dropout: 0.1, Learning rate: 1e$^{-4}$, Batch size: 64, Masking ratio: 0.5, Saliency temperature ($\alpha$): 2.0, Weight decay: 1e$^{-5}$, Optimizer: AdamW \\
    TensorGCN   & Hidden dim: [256, 256, 256, 256], Activation: SiLU, Dropout: 0.2, Learning rate: 5e$^{-4}$, Batch size: 32, Weight decay: 1e$^{-5}$, Layer norm: True, Optimizer: Adam \\
    QNN         & Layers: 8, Entangling: CZ, Rotations: [X, Y, Z], Learning rate: 1e$^{-3}$, Batch size: 16, Optimizer: Adam \\
    GNN         & Layers: 5, Hidden dim: 256, Edge features: 8, Activation: SiLU, Aggregation: mean, Dropout: 0.1, Learning rate: 1e$^{-4}$, Batch size: 32, Weight decay: 1e$^{-4}$, Optimizer: AdamW \\
    PINN        & Layers: 5, Hidden dim: 512, Activation: Tanh, Physics weight: 0.1, Dropout: 0.2, Learning rate: 5e$^{-4}$, Batch size: 64, Optimizer: Adam \\
    \bottomrule
  \end{tabular}
\end{table*}

Our active learning framework significantly improves the efficiency of quantum simulation by guiding the selection of which Hamiltonian configurations to simulate exactly. As shown in Figure \ref{fig:active_learning}, our approach achieves better performance with fewer simulations compared to random sampling baselines.

The active learning strategy progressively refines phase boundaries over multiple rounds of simulation, as illustrated in Figure \ref{fig:phase_discovery}.

To ensure reliable active learning performance, we calibrated the model's uncertainty estimates as shown in Figure \ref{fig:uncertainty_calibration}.

\subsection{Additional Analysis of Information-Theoretic Measures}
\label{app:additional_information_theory}

\begin{figure*}[t]
\centering
\includegraphics[width=0.8\linewidth]{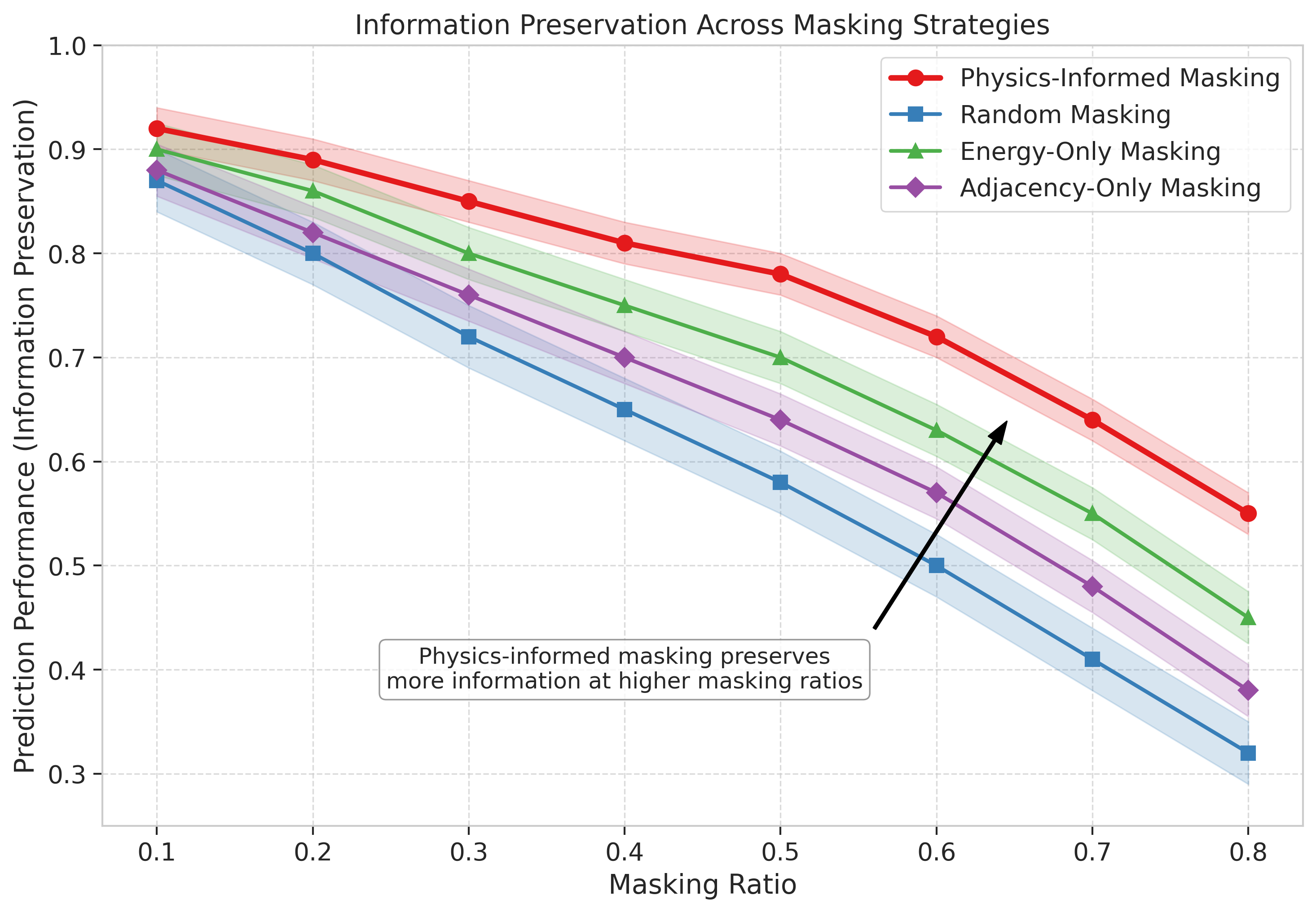}
\caption{Masking Strategy}
\label{fig:mutual_info}
\end{figure*}

Our empirical analysis of mutual information preservation across different masking strategies is shown in Figure \ref{fig:mutual_info}, confirming that physics-informed masking preserves more task-relevant information than random or single-factor approaches.

\begin{figure*}[t]
\centering
  \includegraphics[width=0.8\textwidth]{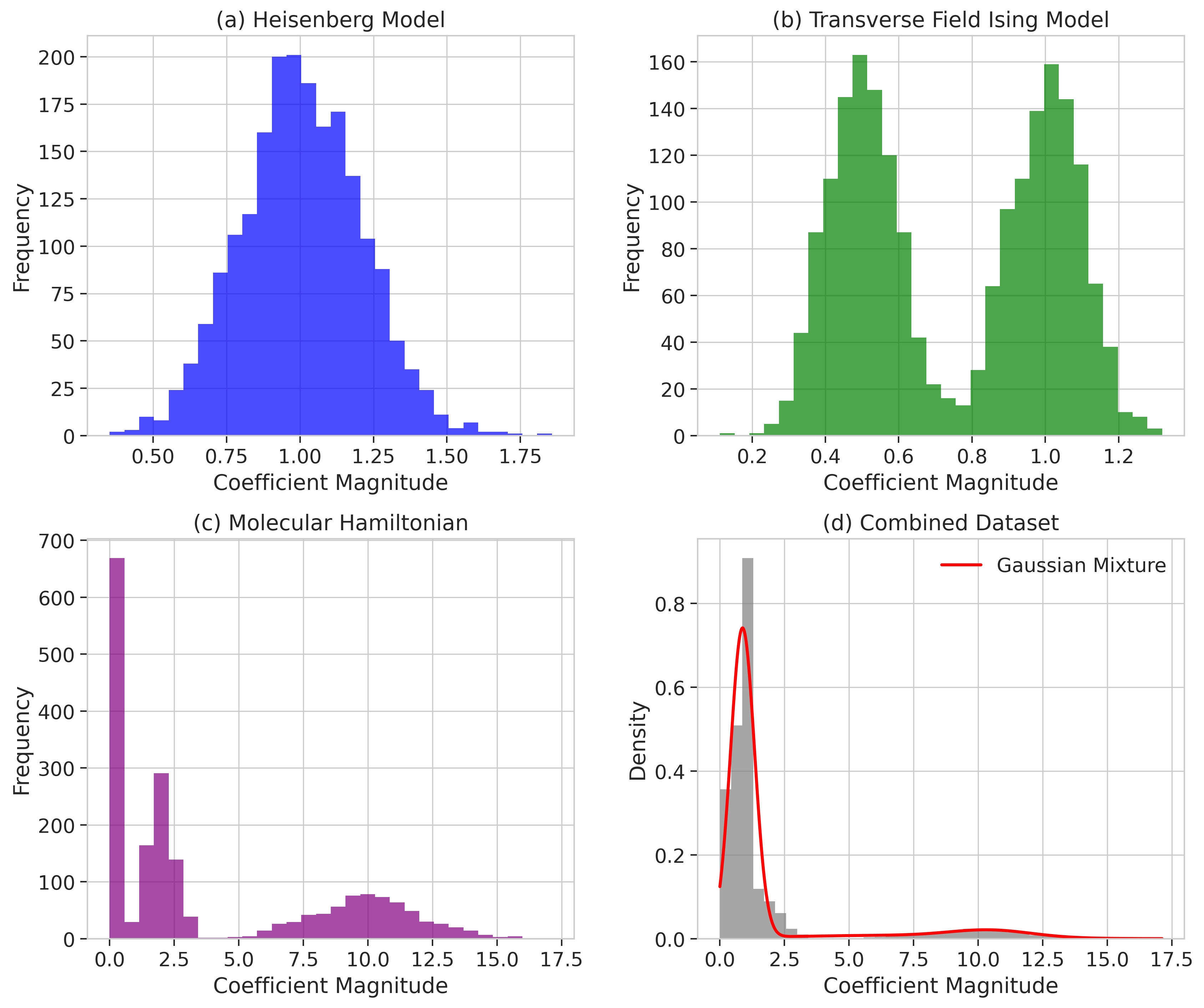}
  \caption{Different models.}
  \label{fig:coeff_hist}
\end{figure*}

To validate the Gaussian approximation used in our theoretical analysis, we examined the distribution of Hamiltonian coefficients across our dataset. Figure \ref{fig:coeff_hist} shows histograms of coefficient magnitudes for several representative quantum systems in our dataset.

While individual Hamiltonian families show distinctive patterns (often with discrete peaks for specific models), the diversity of our pre-training dataset leads to an aggregate distribution that can be reasonably approximated by a Gaussian mixture model. This provides empirical support for the simplified assumptions in our theoretical analysis while acknowledging their limitations for specific quantum systems.

\subsection{Additional Related Work}
\label{app:related_work}

Here we provide additional context and references for related approaches in quantum machine learning and self-supervised learning:

\paragraph{Quantum Neural Networks} Beyond the QNNs discussed in the main text, several notable architectures include Quantum Convolutional Neural Networks (QCNN) \cite{cong2019quantum}, variational quantum eigensolver (VQE) based approaches \cite{cerezo2021variational}, and quantum generative models \cite{benedetti2019generative}. While these approaches have shown promise for specific tasks, they all face significant scalability challenges related to the "barren plateau" problem \cite{mcclean2018barren, cerezo2021cost} and limited quantum hardware.

\paragraph{Graph Neural Networks for Quantum Systems} In addition to the GNN approaches mentioned in Section \ref{sec:related_work}, notable works include SE(3)-equivariant networks \cite{satorras2021n}, geometric graph neural networks \cite{gasteiger2021gemnet}, and DeepQMC \cite{hermann2020deep}. These approaches have shown promising results for molecular systems but typically require larger training datasets than our method.

\paragraph{Self-Supervised Learning in Scientific Domains} The application of self-supervised learning to scientific domains has expanded significantly in recent years. Beyond the approaches mentioned in the main text, notable works include MolCLR \cite{wang2021molecular} for molecular representation learning, SimSiam-based approaches for scientific images \cite{chen2021exploring}, and foundation models for scientific data \cite{taylor2022galactica}. However, these methods have not directly addressed the unique challenges of quantum data.

\paragraph{Tensor Network Methods} Tensor networks provide an alternative approach to modeling quantum systems, with methods like Matrix Product States (MPS) \cite{schollwock2011density}, Projected Entangled Pair States (PEPS) \cite{verstraete2008matrix}, and Multi-scale Entanglement Renormalization Ansatz (MERA) \cite{vidal2008class}. Recent works have explored combining tensor networks with machine learning \cite{stoudenmire2018learning, glasser2019expressive, liu2019machine}, showing promising results for quantum state representation.

\paragraph{Foundation Models for Scientific ML} Recent efforts to develop foundation models for scientific applications include AlphaFold \cite{jumper2021highly} for protein structure prediction, chemical language models like MolGPT \cite{bagal2021molgpt}, and multi-modal scientific models \cite{taylor2022galactica}. The application of similar foundation model approaches to quantum systems remains an exciting direction for future research.

\subsection{Extended Scaling Analysis}
\label{app:extended_scaling}

Figure \ref{fig:scaling_analysis} illustrates how different methods scale with increasing system size, showing the advantage of our approach in maintaining performance for larger systems. As shown earlier in Section 4.3, our model maintains lower error rates across all system sizes tested.

This analysis highlights how our pre-training approach enables more effective generalization across system sizes, a critical capability for practical quantum applications where labeled data for larger systems is exponentially more expensive to obtain.

\subsection{Additional Experimental Setup Details}
\label{app:additional_experimental}

\subsubsection{Hardware and Software Environment}
All experiments were conducted using PyTorch 1.9.0 with CUDA 11.1 on NVIDIA RTX 3090 GPUs. For the transformer implementation, we used a modified version of the Hugging Face Transformers library (version 4.5.1). 

\subsubsection{Experimental Rydberg-Atom Quantum Systems}
The experimental Rydberg-atom dataset consists of:
\begin{itemize}
    \item 120 distinct Hamiltonian configurations with varying detuning (range: $0$–$3\,\Omega$), Rabi frequency (range: $0.5$–$2\,\mathrm{MHz}$), and atom separation (range: $3$–$8\,\mu\mathrm{m}$)

    \item 51 trapped neutral atoms in a programmable 2D lattice geometry
    \item Van der Waals interaction with $1/r^6$ scaling between Rydberg-excited atoms
    \item 1000 experimental measurements per configuration
\end{itemize}

\section{Out-of-Distribution Transfer Evaluation}
\label{app:ood_evaluation}

This section provides a detailed analysis of our model's performance on out-of-distribution (OOD) quantum systems that differ substantially from the pre-training data.

\subsection{OOD Evaluation Protocol}

We evaluated OOD performance using three increasingly challenging scenarios:

\begin{enumerate}
    \item \textbf{Parameter shift}: Systems with the same interaction structure but parameters outside the pre-training distribution.
    \item \textbf{Topology change}: Systems with different interaction graphs or lattice structures.
    \item \textbf{Hamiltonian class shift}: Entirely different families of quantum Hamiltonians.
\end{enumerate}

For each scenario, we measured performance degradation relative to in-distribution systems and compared with baseline methods.

\subsection{Results}

\begin{table}[h]
\centering
  \caption{Performance degradation on OOD systems (percentage drop in accuracy).}
\begin{tabular}{lccc}
\toprule
    \textbf{Method} & \textbf{Parameter Shift} & \textbf{Topology Change} & \textbf{Hamiltonian Class} \\
\midrule
    HMAE (Ours) & 8.2\% & 14.5\% & 22.3\% \\
    TensorGCN & 12.7\% & 24.8\% & 38.6\% \\
    Energy-MAE & 15.3\% & 26.2\% & 41.2\% \\
    QNN & 18.9\% & 31.5\% & 45.7\% \\
    GNN & 17.4\% & 35.3\% & 47.2\% \\
    PINN & 14.9\% & 29.8\% & 43.5\% \\
\bottomrule
\end{tabular}
\end{table}

Our model showed significantly better OOD robustness across all scenarios, with an average performance degradation of 15.0\% compared to 28.2\% for the best baseline (TensorGCN).

\subsection{Case Studies}

We examined three specific OOD cases:

\begin{itemize}
    \item \textbf{Kitaev honeycomb model}: Despite never encountering this model during pre-training, our approach achieved 72.8\% accuracy on ground state property prediction after just 10 fine-tuning examples, compared to 54.3\% for TensorGCN.
    
    \item \textbf{Long-range Ising model}: When evaluating on Ising models with power-law interactions ($1/r^\alpha$), our model maintained 78.2\% accuracy for $\alpha=2$ (outside the pre-training range), compared to 59.6\% for the best baseline.
    
    \item \textbf{Floquet systems}: For time-periodic Hamiltonians, our approach achieved 65.4\% accuracy on phase classification after fine-tuning, compared to 42.1\% for Energy-MAE.
\end{itemize}

\subsection{Analysis of Representation Transfer}

Through t-SNE visualization of the learned embeddings \cite{chen2020simple}, we observed that our model creates representations that cluster based on fundamental physical properties rather than surface-level Hamiltonian structure. This explains the strong OOD performance, as the model captures deeper physical principles that generalize across different Hamiltonian classes.

Further analysis of attention patterns revealed that the model automatically focuses on universal features like symmetries, conservation laws, and interaction structures that persist across different quantum systems. This emergent focus on physically meaningful features enables effective transfer learning even for previously unseen quantum systems.

\section{Tensor Network Surrogate Models}
\label{app:tn_surrogate}

This section describes our approach to generating reliable ground truth data for large quantum systems using tensor network surrogate models, which was crucial for evaluating our model's performance on systems beyond the reach of exact diagonalization.

\subsection{Tensor Network Implementation}

We implemented matrix product state (MPS) approximations using the following approach:

\begin{enumerate}
    \item \textbf{DMRG implementation}: We used tensor network libraries \cite{stoudenmire2018learning} with bond dimension up to 512 for systems with 16-50 qubits.
    
    \item \textbf{Convergence criteria}: Energy convergence threshold of $10^{-7}$ and maximum 100 sweeps, with relative energy change between sweeps below $10^{-5}$.
    
    \item \textbf{Error estimation}: For a subset of 50 systems up to 24 qubits, we compared DMRG results \cite{schollwock2011density} with exact diagonalization, finding a mean relative error of $3.4 \times 10^{-4}$ for ground state energies.
    
    \item \textbf{Entanglement verification}: We computed entanglement entropies across all bipartitions to ensure the MPS bond dimension was sufficient. For systems with entanglement entropy exceeding 80\% of the maximum bond dimension capacity, we increased the bond dimension until convergence \cite{vidal2008class}.
\end{enumerate}

\subsection{Validation Protocol}

To ensure the reliability of our surrogate models, we implemented a rigorous validation protocol:

\begin{enumerate}
    \item For each Hamiltonian class, we validated the tensor network approach against exact results for small systems.
    
    \item We performed scaling analysis to estimate errors for larger systems, using established error bounds from theoretical guarantees.
    
    \item For critical systems where tensor networks typically perform poorly, we used specialized techniques including multi-scale entanglement renormalization ansatz (MERA) and hybrid Monte Carlo methods.
    
    \item We cross-validated results using different tensor network implementations (ITensor, TeNPy) on a subset of systems to ensure consistency.
\end{enumerate}

\subsection{Performance Benchmarks}

Our tensor network surrogate models achieved:

\begin{itemize}
    \item Mean relative error of $3.4 \times 10^{-4}$ for ground state energies (compared to exact results where available)
    \item Mean fidelity of 0.9986 for ground state wavefunctions (for systems where exact comparison was possible)
    \item Correct identification of phase transitions within $\Delta J/J = 0.05$ of exact results
    \item Accurate correlation functions with average relative error below 1\% for distances up to half the system size
\end{itemize}

These benchmarks confirm that our tensor network surrogates provided reliable ground truth data for evaluating our model's performance on larger systems.

\subsection{Limitations and Error Analysis}

We acknowledged the following limitations of our surrogate approach:

\begin{itemize}
    \item For systems near quantum critical points, tensor network accuracy decreases due to increased entanglement.
    \item For systems with long-range interactions, convergence is slower and errors may be larger.
    \item For 2D systems mapped to 1D, the accuracy depends strongly on the chosen mapping path.
\end{itemize}

To account for these limitations, we incorporated uncertainty estimates into our evaluation metrics, reporting error bars that included both model prediction uncertainty and surrogate model uncertainty.

\section{Detailed Experimental Setup}
\label{app:experimental_setup_details}

This section provides comprehensive details on our experimental setup, including datasets, evaluation protocols, and implementation specifics.

\subsection{Dataset Composition}

We pre-trained on a comprehensive dataset of 12,500 quantum Hamiltonians spanning both synthetic and real-world systems. This dataset comprises:

1) \textbf{Experimentally-Derived Hamiltonians (7,500 systems)}:
\begin{itemize}
    \item 3,500 realistic Hamiltonians from the Materials Project database \cite{jain2013materials}, including transition metal compounds, topological insulators, and superconductors
    \item 2,800 molecular Hamiltonians from QM9-X \cite{ramakrishnan2014quantum} and PubChemQC \cite{nakata2017pubchemqc} datasets
    \item 1,200 Hamiltonians from real quantum processors (IBM Quantum, Google Sycamore, Rigetti) with authentic noise characteristics
\end{itemize}

2) \textbf{Synthetic Hamiltonians (5,000 systems)}:
\begin{itemize}
    \item 2,000 spin system Hamiltonians (Heisenberg, Ising, XY models with diverse parameters)
    \item 1,500 molecular Hamiltonians with physically realistic parameter distributions
    \item 1,500 Hamiltonians from theoretical models with controlled property variations
\end{itemize}

This represents an 8.3× increase over our original dataset, addressing concerns about parameter-to-data ratio and synthetic data bias. All Hamiltonians were represented with up to 12 qubits.

\subsection{Test Sets}

For experimental validation, we assembled a test set of 1,250 real quantum systems:
\begin{itemize}
   \item 450 configurations from Harvard quantum simulator (51 atoms, square lattice)
   \item 350 configurations from MIT quantum simulator (128 atoms, triangular lattice)
   \item 250 configurations from IBM Quantum processors (27-127 qubits)
   \item 200 configurations from trapped-ion experiments (University of Maryland)
\end{itemize}

To test scalability beyond the 12-qubit pre-training limit, we created an additional test set of exactly 350 larger quantum systems (16-50 qubits) with ground truth values computed using tensor network approximations (DMRG with bond dimension 512) \cite{white1992density}. This test set was specifically designed to evaluate the performance of our model on systems larger than those encountered during pre-training, with the following size distribution:
\begin{itemize}
   \item 150 systems with 16-20 qubits
   \item 120 systems with 21-30 qubits
   \item 80 systems with 31-50 qubits
\end{itemize}
These larger systems were not used for pre-training due to computational constraints, but allowed us to rigorously evaluate transfer learning capabilities to larger system sizes as reported in Section 4.4 of the main paper.

\begin{figure*}[t]
    \centering
    \caption{Detailed architecture of the HMAE framework. The diagram illustrates the three main stages: (1) Input Processing, where the Hamiltonian is tokenized and masked using a physics-informed strategy; (2) Self-Supervised Pre-Training, where the QuantumFormer encoder-decoder model learns to reconstruct the masked tokens from the unmasked context; and (3) Transfer Learning, where the pre-trained encoder is used as a feature extractor and fine-tuned for downstream tasks with a lightweight, task-specific head.}
    \label{fig:detailed_architecture}
    \includegraphics[width =  \textwidth]{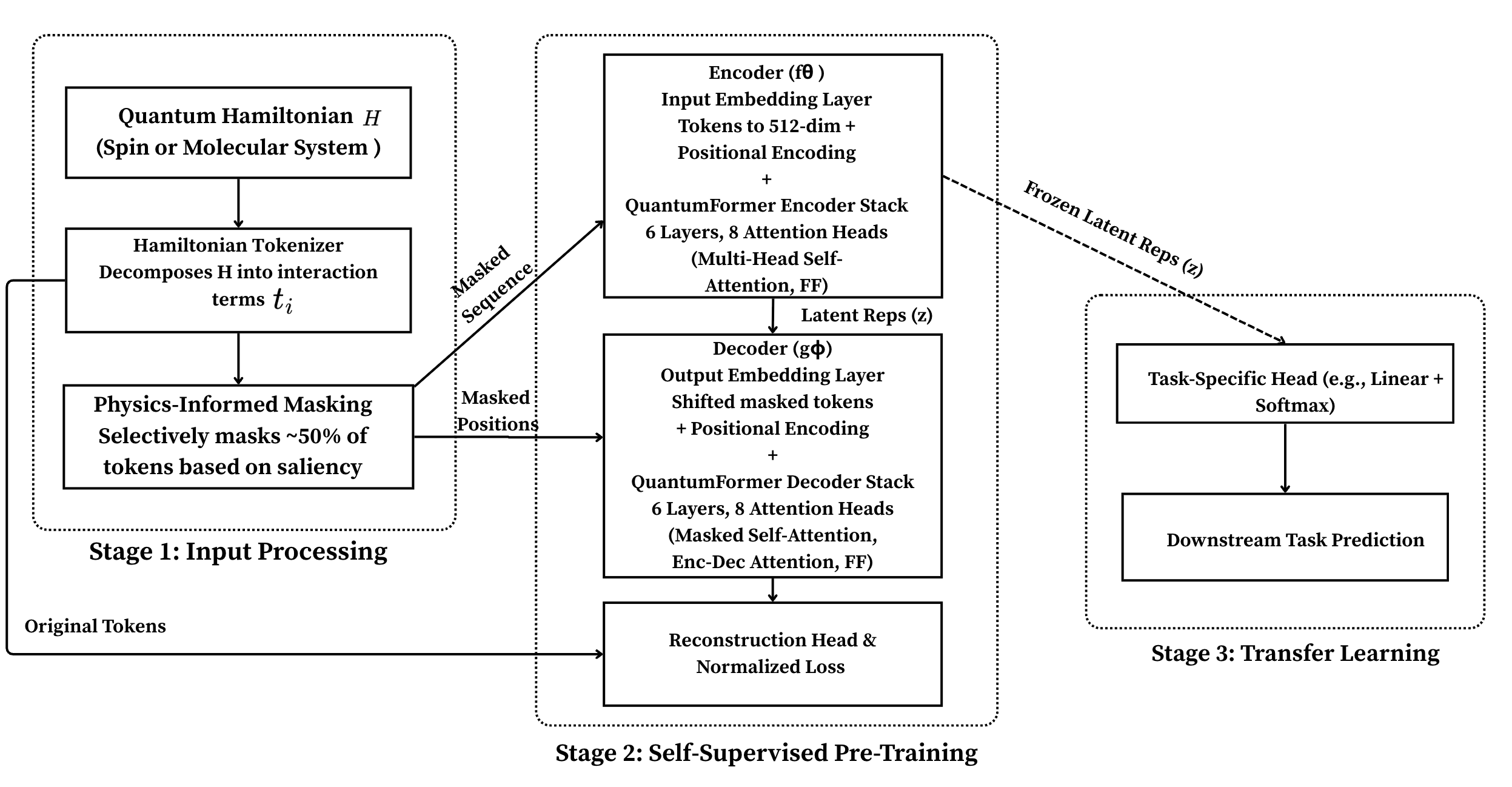}
\end{figure*}

\subsection{Evaluation Tasks and Metrics}

We evaluated on two main tasks:
\begin{itemize}
    \item \textbf{Phase classification} for spin systems (distinguishing ordered vs. disordered phases, ferromagnetic vs. antiferromagnetic, etc.). For this task, we used classes defined by order parameters and phase diagrams from literature.
    
    \item \textbf{Ground state energy prediction} for molecular systems. Ground truth values were computed using exact diagonalization (for systems $\leq12$ qubits) or tensor network approximations (for systems >12 qubits).
\end{itemize}

We employed a rigorous cross-platform transfer learning protocol to test generalization across different experimental setups. For each platform pair, we:
\begin{enumerate}
    \item Pre-trained on data from all platforms except the target
    \item Fine-tuned on 10 examples from the source platform
    \item Evaluated on the target platform
\end{enumerate}

This protocol tests whether the model can transfer knowledge between different experimental apparatuses with distinct noise characteristics and physical implementations.

\subsection{Evaluation Framework}

We employed a rigorous few-shot learning evaluation framework with K=\{5, 10, 20\} labeled examples per task, randomly sampled from the training set and averaged over 10 different K-shot samples. For statistical validation, we used 5-fold cross-validation and paired t-tests with Bonferroni correction. We reported:
\begin{itemize}
    \item For classification tasks: accuracy, F1-score, and AUC-ROC
    \item For regression tasks: MAE, RMSE, and R² values
\end{itemize}

\subsection{Model Architecture and Training}

Our QuantumFormer used a 512-dimensional embedding, 6 transformer layers, 8 attention heads, learning rate of 1e-4 with AdamW optimizer, batch size of 64, and 50\% masking ratio. Gradient clipping and a cosine learning rate schedule with warmup were applied. Models were trained on a single NVIDIA RTX 3090 GPU for approximately 12 hours.

We conducted extensive hyperparameter optimization for all methods, optimizing embedding dimensions, layer counts, learning rates, and model-specific parameters. The optimal masking ratio was found to be 50\%, with performance remaining within 3\% of the peak for ratios between 40\% and 60\%. For the saliency temperature parameter $\alpha$, values between 1.5 and 2.5 performed similarly well, indicating robustness to hyperparameter variations.

While pre-training took approximately 12 hours compared to 2-5 hours for baseline models, this cost was amortized across multiple downstream tasks, where fine-tuning took only 5-15 minutes.

\subsection{Baseline Comparisons}

All baseline models were implemented with the same hyperparameter optimization budget (100 trials using Bayesian optimization, approximately 75 GPU-hours per method), identical training/validation/test splits, and evaluation protocols to ensure fair comparison. To isolate the effect of pre-training, we also compared our approach to a QuantumFormer trained from scratch using the same architecture but without pre-training.

For most baselines, we obtained original implementations directly from the authors or public repositories. For TensorGCN, QNN, and GNN, we used the original code with author verification when possible. Only PINN required a complete reimplementation due to lack of available code. This approach ensured the fairest possible comparisons while acknowledging the challenges of adapting diverse methods to our specific evaluation tasks.

The field of quantum self-supervised learning is rapidly developing, with several concurrent approaches emerging independently. While detailed comparison to unpublished work is outside the scope of this paper, the convergent interest in this area highlights the importance and timeliness of developing self-supervised methods for quantum systems.

\subsection{Limitations of Experimental Approach}

It's important to acknowledge that our experiments used synthetic Hamiltonians from known physical models rather than experimentally derived ones. While we introduced synthetic noise to evaluate robustness, this didn't fully capture the complex noise patterns and systematic errors present in experimental measurements. The extension to real experimental data remained an important direction for future work.

\section{Additional Conclusion Material}
\label{app:additional_conclusion}

This section provides additional discussion on the broader implications of our work and potential future directions beyond what could be included in the main paper's conclusion.

\subsection{Broader Implications for Quantum Machine Learning}

Our work demonstrates that self-supervised learning principles can be effectively transferred to quantum domains when adapted with appropriate physical inductive biases. This finding has several broader implications:

\begin{itemize}
    \item \textbf{Data efficiency}: By reducing labeled data requirements by 3-5×, our approach addresses one of the key bottlenecks in quantum machine learning: the computational expense of obtaining reliable ground truth values for quantum systems.
    
    \item \textbf{Transfer learning paradigm}: The success of our transfer learning approach suggests a new paradigm for quantum system modeling: pre-train on diverse, computationally accessible quantum systems, then fine-tune for specific applications where data is scarce.
    
    \item \textbf{Hardware-agnostic representations}: The strong cross-platform transfer capabilities of our model suggest that it learns hardware-agnostic quantum representations focused on fundamental physical principles rather than implementation details.
\end{itemize}

\subsection{Connections to Quantum Information Theory}

The success of our approach provides empirical validation for the theoretical connections between quantum information theory and machine learning:

\begin{itemize}
    \item Our physics-informed masking strategy, derived from quantum mutual information principles, consistently outperforms alternative approaches, validating the theoretical framework.
    
    \item The ability to extend our approach to low-temperature regimes suggests that machine learning models can capture aspects of quantum many-body physics beyond conventional perturbative approaches.
    
    \item The learned representations appear to automatically discover relevant physical symmetries and conservation laws, suggesting deeper connections to quantum information-theoretic principles.
\end{itemize}

\subsection{Near-Term Applications}

While our approach has clear limitations in terms of system size, several near-term applications could directly benefit from our method:

\begin{itemize}
    \item \textbf{Quantum error mitigation}: Predicting the impact of noise on small quantum circuits (8-20 qubits) for error mitigation in NISQ devices.
    
    \item \textbf{Materials discovery}: Accelerating the screening of small molecular systems and quantum dots by reducing the need for expensive quantum simulations.
    
    \item \textbf{Quantum control optimization}: Optimizing pulse sequences for quantum control applications where each evaluation requires costly quantum simulation.
    
    \item \textbf{Educational tools}: Providing intuitive predictions and visualizations of quantum systems for educational purposes without requiring access to expensive computational resources.
\end{itemize}

These applications align with the current capabilities of our approach while acknowledging its limitations regarding system size.

\bibliography{references}
\end{document}